\documentclass[journal,12pt,onecolumn,draftcls,twoside]{IEEEtran}
\normalsize
\usepackage{verbatim}
\usepackage{algorithmicx}
\usepackage{algpseudocode}
\usepackage{algorithm}
\usepackage{hyperref}
\usepackage[caption=false,font=footnotesize]{subfig}
\usepackage[pdftex]{graphicx}
\usepackage{enumitem}
\usepackage{amsmath,amssymb}
\usepackage{amsthm}
\usepackage{cite}
\usepackage{stfloats}
\theoremstyle{plain}
\newtheorem{theorem}{Theorem}
\newtheorem{lemma}{Lemma}
\newtheorem{claim}{Claim}
\newtheorem{corollary}{Corollary}
\theoremstyle{definition}
\newtheorem{definition}{Definition}
\newtheorem{example}{Example}
\usepackage{enumitem}
\theoremstyle{remark}

\newtheorem{case}{Case}
\newcommand{\floor}[1]{\lfloor #1 \rfloor}
\newcommand{\ceil}[1]{\lceil #1 \rceil}
\usepackage{array}
\usepackage{tabularx}
\usepackage{ragged2e} 
\usepackage{url}
\allowdisplaybreaks
\begin{document}
\title{Shared Cache Coded Caching Schemes Using Designs and Circuits of Matrices}
\author{Niladri Das and
        B. Sundar Rajan, \textit{Fellow}, \textit{IEEE}
\thanks{Manuscript received on date xx; revised on date xx; accepted on date xx. Date of publication xx,; date of current version xx. The editor coordinating the review of this paper and approving it for publication was Prof. xx.}
\thanks{Niladri Das and B. Sundar Rajan are with the Department
of Electrical Communication Engineering, Indian Institute of Science, Bangalore, 
India e-mail: niladribegins@gmail.com and bsrajan@iisc.ac.in.}
%
\thanks{This work was supported partly by the Science and Engineering Research Board (SERB) of the Department of Science and Technology (DST), Government of India, through J.C. Bose National Fellowship to B. Sundar Rajan and SERB National Post Doctoral Fellowship (No. PDF/2019/000044) to Niladri Das.}
\thanks{Part of this paper has been submitted to the 2023 IEEE Information Theory Workshop (ITW) to be held in France.}}

\maketitle

\begin{abstract}
In this paper, we study shared cache coded caching (SC-CC): a set of caches serves a larger set of users; each user access one cache, and a cache may serve many users. For this problem, under uncoded placement, Parrinello, Ünsal, and Elia showed an optimal SC-CC scheme, in which the subpacketization level depends upon the number of caches. We show an SC-CC scheme where the subpacketization level does not directly depend upon the number of users or caches; any number of caches and users can be accommodated for a fixed subpacketization level. Furthermore, new caches can be added without re-doing the placement of the existing caches. We show that given an upper limit on the allowable subpacketization level, our SC-CC scheme may achieve a lesser rate than other relevant SC-CC schemes. Our scheme is constructed using matrices and  designs. A matroid can be obtained from a matrix over a finite field; the placement of our scheme is decided by a design constructed from a matrix; the circuits of a matroid obtained from the matrix and the design is used to decide the delivery. 
\end{abstract}
%
%
%
%
%
%

\IEEEpeerreviewmaketitle

\section{Introduction}
\IEEEPARstart{I}{n} the past decade, the world has witnessed an ever-increasing demand for content download over the wireless medium. Our subject for this paper is content delivery to a set of users connected to a central server (like a base station) over a shared error-free broadcast channel. The idea behind caching is to make content available closer to the users so that at least a part of a user's demand can be downloaded directly from the local cache. This helps alleviate the load on the available bandwidth of the shared broadcast channel between the users and the base station.  

The concept of coded caching was introduced in the seminal work by Maddah-Ali and Niesen in \cite{ali}. It can be seen that because of the caches, the server needs to broadcast fewer bits than if there had been no caches. However, the reduction is not limited to the storage capacity of the individual caches. In \cite{ali} Maddah-Ali and Niesen showed that a multicasting gain (called coding gain) could be obtained by carefully placing content in the caches and then broadcasting coded packets.

The model considered in \cite{ali} assumes that each cache is accessed by only one user -- we call this problem the dedicated cache coded caching (DC-CC) problem. In a practical scenario, it seems likely that several users would access the same cache. For example, a content caching architecture (called Femtocaching) over the wireless cellular network was  proposed by Shanmugam, Golrezaei, Dimakis, Molisch and Caire in \cite{femto}; in this architecture, small-cell base stations act as the caches, and the macro-cell base station acts as the central server. Naturally, the small-cell base stations would serve several users. Another interesting architecture proposed by Wan, Tuninetti, Ji and Caire in \cite{tuni} has SC-CC as an integral part. Zhao, Bazco-Nogueras, and  Elia showed  an added advantage: the near-far effect can be mitigated by deploying a shared cache setup \cite{s:near-far}.

\subsection{Related work}
The SC-CC problem was first studied by Maddah-Ali and Niesen in \cite{ali2}. The authors considered the case when each cache serves an equal number of users. Xu, Gong, and Wang showed an SC-CC scheme with coded placement given small cache sizes \cite{s:what4}. Parrinello, Ünsal, and Elia constructed an SC-CC scheme and showed that their scheme is optimal under uncoded placement \cite{elia}. The scheme in \cite{elia} is built upon the DC-CC scheme shown in \cite{ali}, which in turn has been shown to be optimal under uncoded placement in \cite{kaiwan,Yu}. Dutta and Thomas showed a decentralized scheme for the SC-CC problem in \cite{dutta}. Recently, Peter and Rajan in \cite{peter} used Placement Delivery Arrays (a PDA is a structure introduced by Yan \textit{et al.} in \cite{yctc}) to construct an SC-CC scheme. The authors modified existing PDAs to construct a new PDA suitable for SC-CC. The resulting SC-CC scheme transfers the benefits of lower subpacketization level offered by PDA based DC-CC schemes to the shared cache setup. The authors also showed that the scheme in \cite{elia} is a special case of their scheme (constructed from the MN PDA shown in \cite{yctc}). Works in \cite{elia, peter, lamp} study SC-CC when the broadcaster is equipped with multiple antennas. 

For the case when the server knows the number of users accessing each cache before placement, Ibrahim, Zewail, and Yener showed a scheme with coded placement in \cite{s:what}. Peter, Namboodiri, and Rajan also addressed the same case and leveraged the known user distribution to modify PDAs accordingly to achieve better results. Reference \cite{s:elia2} studies memory size allocation of the caches  based upon the number of users accessing each cache. 

The multiple file request problem refers to a version of the DC-CC problem where each user demands multiple files. An equivalence between this problem and the SC-CC problem has been shown in \cite{elia}. Interesting works on multiple file requests (when each user requests the same number of files) can be found in references \cite{Ji2, sengupta, zhang-wang, chandani, ulukus}.

\subsection{Using DC-CC schemes to construct SC-CC schemes}

In SC-CC, each cache is accessed by a set of users, and each user accesses one cache. An approach towards finding an SC-CC scheme is to break the SC-CC problem into several DC-CC sub-problems, then use a TDMA-like approach and solve each DC-CC sub-problem in each time division. For example, say there are three caches $C_1,C_2,C_3$, and users $u_1, u_2, u_3$ access $C_1$, users $u_4, u_5$ access $C_2$, user $u_6$ accesses $C_3$. In the first DC-CC sub-problem $u_1$ accesses $C_1$, $u_4$ accesses $C_2$, and $u_6$ accesses $C_3$; in the second DC-CC sub-problem $u_2$ accesses $C_1$, and $u_5$ accesses $C_2$;  in the third DC-CC sub-problem $u_3$ accesses $C_1$. Each of these DC-CC sub-problems is addressed separately in each time division using a DC-CC scheme.  In fact, the SC-CC schemes in \cite{elia} and \cite{peter}, to the best of our understanding, at its foundational level, use the same strategy. Our scheme is also built upon the same idea; roughly, the difference is in how the SC-CC problem is split into several DC-CC sub-problems and in the DC-CC scheme used to solve each sub-problem.
%
\subsection{Subpacketization bottleneck}
The scheme shown in \cite{ali} requires each file to be split into a number of equal-sized disjoint fragments or subfiles. This number is called the subpakcetization level. The file size (in bits) has to be larger than the number of subfiles for it to be fragmented. A high subpacketization level also indicates more computations at the users' devices, which generally have limited power resources. The subpacketization level of the scheme in \cite{ali} increases exponentially with the number of users. Reducing the subpacketization level, however, may increase the number of bits the server has to broadcast to fulfill the demands of the users \cite{shanmugam, yctc}. The problem of high subpacketization is referred to as subpacketization bottleneck.

Several techniques have been used in the literature to address the problem of high subpacketization level. Some of these techniques are: grouping scheme in \cite{shanmugam}; PDAs have been used in many papers such as \cite{yctc,flexible_memory,variant,framework,hypergraph,bipartite_graph,Linear_Subpacketization}; Ruzsa-Szeméredi Graphs \cite{rs_graph,other2}; a combinatorial structure called resolvable design \cite{rama}; projective geometry \cite{krishnan1,chittoor1}; other combinatorial designs \cite{li,hypercube,other2}; multiple antennas at the central server \cite{adding_transmitters,other6}.

\subsection{Problem statement}
\begin{figure}[htpb]
\centering
\includegraphics[width=0.9\textwidth]{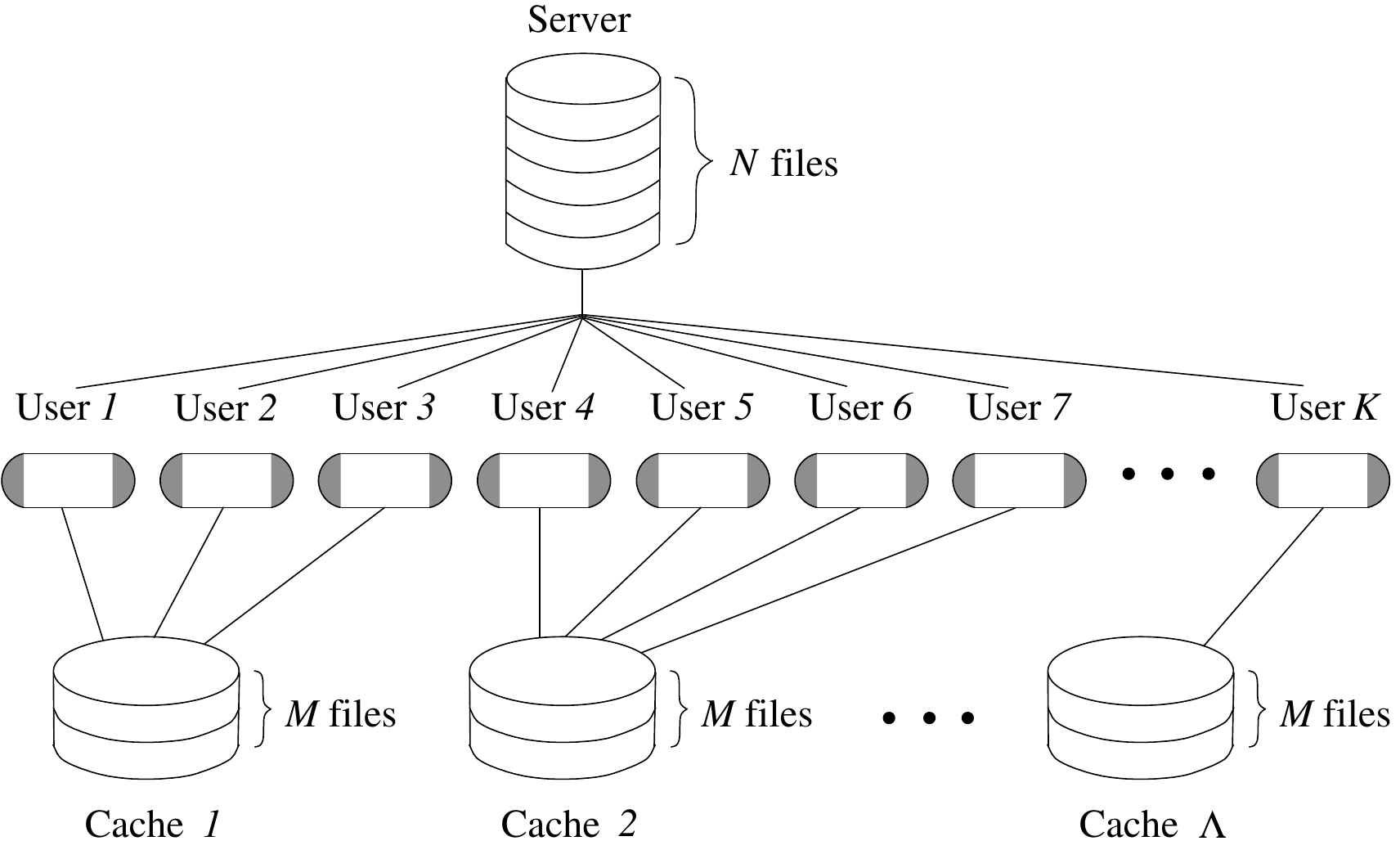}
\caption{A pictorial description of an example SC-CC problem. The $K$ users are connected to the central server through a shared error-free link. Each user accesses one of the caches. Cache $1$ is accessed by User $i$ for $i=1,2,3$; Cache $2$ is accessed by User $i$ for $i=4,5,6,7$; User $K$ accesses Cache $\Lambda$. If Cache $i$ is denoted by $c_i$ and $L_i$ denotes the number of users that accesses the cache $c_i$, then we have $L_1 = 3$, $L_2 = 4$, $L_{\Lambda}\geq 1$.}
\label{graph1}
\end{figure}
Our problem has (i) a central server with $N$ files, (ii) $K$ users, each user demands one of the $N$ files at the central server, and (iii) $\Lambda$ caches, each cache has a storage capacity of bits equivalent to $M$ files.  Each user accesses one of the $\Lambda$ caches, and a user can instantaneously download the contents of the cache it accesses. 

If each cache is accessed by exactly one user, the problem is called the dedicated cache coded caching (DC-CC). Note,  DC-CC implies $\Lambda = K$. If there exists at least one cache which is accessed by more than one user, the problem is called the shared cache coded caching (SC-CC). 

User-to-cache association is a list specifying which user accesses which caches. For the SC-CC problem, given the  $\Lambda$ caches are denoted by $c_1,c_2,\ldots,c_\Lambda$, the user-to-cache association profile $\mathcal{L} = \{L_i\,|\, 1\leq i\leq \Lambda\}$ is a set of $\Lambda$ integers where $L_i$ denotes the number of users that accesses the cache $c_i$. An example SC-CC problem is shown in Fig.~\ref{graph1}. 

The central server first places bits of  size equivalent to $M$ files in each of the caches without knowing the user-to-cache association (this is called the placement phase). Then, each of the $K$ users chooses one of the caches to access. Each user then communicates the following to the central server: (i) the cache it accesses and (ii) the file it demands. Next, the central server broadcasts bits of size equivalent to $R$ files  (this is called the delivery phase). Each user retrieves the file it demands using the broadcast and the contents of the cache it accesses. The objective of the problem is to minimize $R$. It is to be noted that the central server also informs all users of each other's demands, and the bits required for this communication are not accounted for in $R$ files.

It is likely that in a practical scenario $N >> K$ and the demands of the users are distinct. Thus, the refined objective is to minimize $R$ without assuming any intersection between the demands of the users (a scenario that is the worst case; if two users demand the same file, it may be possible to exploit the intersection of demands and reduce $R$ further).

A DC-CC / SC-CC scheme specifies: (i) the bits (of size equivalent to $M$ files) to place in each cache, (ii) the bits (of  size equivalent to $R$ files) to broadcast (by the central server) for any possible set of demands. A DC-CC / SC-CC scheme is called uncoded if the entire content of a cache is comprised of segments of the $N$ files at the server. 

A DC-CC / SC-CC scheme may require each file to be split into a certain number of subfiles, and this number is called the subpacketization level. Our objective is to minimize $R$ assuming distinct demands given an upper limit on the subpacketization level.

\subsection{Our Contribution}
%
Consider an SC-CC problem with $\Lambda$ caches, $\frac{M}{N} = \frac{t}{q}$, $t \in \{0,1,\ldots,q\}$, $q$ is a power of prime.
\begin{enumerate}
\item We present an SC-CC scheme for the problem. 
\item The subpacketization level of our SC-CC scheme does not directly depend upon the number of users or caches.  For any $q^m$ (where $m$ is a positive integer) such that $2 \leq m \leq \ceil{\frac{\Lambda}{q}}-1$, our scheme provides a solution having a subpacketization level of $q^m$. This provides a lot of choices to the system administrator in terms of choosing the subpacketization level, which may depend upon external factors such as the underlying communication protocol.
\item In a practical scenario, a cache may be offline during the placement phase due to various reasons, such as a power outage. In our scheme, new caches can be added anytime without requiring any changes to the placement of the existing caches.
\item Our scheme is developed using circuits of matrices (the definition of circuits is borrowed from matroid theory) and a combinatorial structure called designs. 
\item In comparison to the SC-CC scheme by Parrinello \textit{et al.} in \cite{elia}, our scheme has a lesser subpacketization level. In comparison to the reduced subpacketization SC-CC scheme by Peter \textit{et al.} in \cite{peter}, our scheme (i) in some cases achieves a lesser rate for the same subpcketization level and (ii) in some cases achieves the same rate but using a lesser subpcketization level.  
\end{enumerate}

\subsection{Organization of the paper}
In Section~\ref{notations}, we describe the notations used. Section~\ref{pre} is on the mathematical background and preliminaries; we define circuits of matrices and then define designs. In Section~\ref{construct}, we use an existing method to construct designs from matrices and show some properties of the resulting design. In Section~\ref{scheme}, we show the main contribution of the paper: an SC-CC scheme. In Section~\ref{disturb}, we show how new caches can be added without disturbing the placement of the existing caches. In Section~\ref{examples}, we illustrate an example of our SC-CC scheme. In Section~\ref{compare}, we compare our scheme with existing relevant schemes. The paper is concluded in Section~\ref{conclusion}, along with some directions for future work.

\section{Notations}\label{notations}

The notation $(a)_q$ stands for $a \mod q$. Let $G$ be an $k \times m$ matrix over some finite field $\mathbb{F}_q$. The rank of $G$ is denoted by $rank(G)$. The null space of $G$ is denoted by $N(G)$. The content at the $i^{\text{th}}$ row and  $j^{\text{th}}$ column of $G$ is denoted by $G[i,j]$ for $1\leq i\leq k$, $1\leq j\leq m$. The number of rows in $G$ is denoted by $size(G,1)$ (so $size(G,1) = k$). Let the $i^{\text{th}}$ row of $G$ be denoted by $G[i,:]$. Notation $G^T$ indicates the transpose of $G$.

Given $G_i$ is a $k_i \times m$ matrix for $1\leq i\leq n$, let $[G_1 ; G_2 ; \ldots ; G_n]$ denote a matrix of size $(k_1+k_2+\cdots+k_n) \times m$ formed by vertically concatenating  the matrices $G_1,G_2,\ldots,G_n$ in that order.

For any set $\Omega \subseteq \{1,2,\ldots,k\}$, let $G[\Omega,:]$ indicate the matrix formed by vertically concatenating the rows $G[i,:]$ for $\forall i \in \Omega$; that is, if $\Omega =  \{i_1,i_2,\ldots,i_{|\Omega|}\}$ where $1 \leq i_1,i_2,\ldots,i_{|\Omega|}  \leq k$, then $G[\Omega,:] = [G[i_1,:] ; G[i_2,:] ; \cdots ; G[i_{|\Omega|},:]]$.

If $v$ is a row vector, then $v[n]$ indicates the content at the $n^{\text{th}}$ column of $v$. The counting of column or row indices starts from $1$. If $V$ is a set of column vectors and $v$ is a column vector (all vectors are over the same finite field), then $\{V + v\}$ indicates the set of $|V|$ vectors obtained summing $v$ with every vector in $V$. 

In this paper, line x.y indicates line number x of \textbf{Algorithm y} where x and y are numerals.

\section{Mathematical background and preliminaries}\label{pre} 
\subsection{Circuits of a matrix}
Our goal in this subsection is to define circuits of matrices. We have borrowed the terminology ``circuits'' from matroid theory. 
\begin{definition}[Section 1.1 of \cite{oxley}]
A matroid is an ordered tuple $(\mathcal{E},\mathcal{I})$ where $\mathcal{E}$ is a finite set and $\mathcal{I} \subseteq 2^{\mathcal{E}}$ is a set of subsets of $\mathcal{E}$ such that
\begin{description}
\item (i) The empty set is in $\mathcal{I}$.
\item (ii) If $A \subseteq B$ and $B \in \mathcal{I}$, then $A \in \mathcal{I}$.
\item (iii) If $A, B \in \mathcal{I}$ and $|A| < |B|$, then there exists an element $e \in B \setminus A$ such that $A \cup \{e\} \in \mathcal{I}$.
\end{description}
\end{definition}
The set $\mathcal{E}$ is called the ground set, and the elements of $\mathcal{I}$ are called the independent sets. If a subset of $\mathcal{E}$ is not in $\mathcal{I}$, then it is called a dependent set. 
\begin{definition}[Section 1.1 of \cite{oxley}]
A set $C \subseteq \mathcal{E}$ is a circuit of the matroid $(\mathcal{E},\mathcal{I})$ if 
\begin{description}
\item (i) $C \notin \mathcal{I}$.
\item (ii) For any $e \in C$, $C \setminus \{e\} \in \mathcal{I}$.
\end{description}
\end{definition}
Circuits of a matroid can be seen as the minimal dependent sets of the matroid.

We now use a slight variation of the standard way of obtaining a matroid from a matrix. Let $M$ be an $n \times m$ matrix over $\mathbb{F}_q$. Let $\mathcal{E}_M$ be the set of row indices $\{1,2,\ldots,n\}$ and let $\mathcal{I}_M$ be all subsets of $\mathcal{E}_M$ such that, except for the empty set, the rows of $M$ corresponding to the row indices in the subset are independent in the $m$-dimensional vector space over $\mathbb{F}_q$ (the standard way is to consider the set of column indices as the elements of $\mathcal{E}_M$). It can be shown that $(\mathcal{E}_M,\mathcal{I}_M)$ is a matroid (see Proposition 1.1.1 of \cite{oxley} for proof).

In a somewhat abuse of terminology, we refer to the circuits of $(\mathcal{E}_M,\mathcal{I}_M)$ as the circuits of $M$. So,  a set $C \subseteq \{1,2,\ldots,n\}$ is a circuit of $M$ if (i) $rank(M[C,:]) < |C|$ (that is, the set of rows indexed by the elements in $C$ are dependent), (ii) for any $e \in C$, $rank(M[C\setminus\{e\},:]) = |C| - 1$ (that is, the set of rows indexed by the elements in $C\setminus\{e\}$ are independent for any $e \in C$). If a circuit $C$ has $l$ elements, that is $|C| = l$, then $C$ is called an $l$-length circuit. 


\begin{example}
\begin{IEEEeqnarray}{l}
M = \begin{bmatrix}
1 & 0 & 0\\
0 & 1 & 0\\
0 & 0 & 1\\
1 & 1 & 1\\
2 & 1 & 1
\end{bmatrix}
%
\end{IEEEeqnarray}
We have $n = 5, m = 3$. The circuits of $M$ are $\{1,2,3,4\}$, $\{1,2,3,5\}$, $\{1,4,5\}$, $\{2,3,4,5\}$. Set $\{1,3,4\}$ is not a circuit because the corresponding columns of $M^T$ are independent. Set $\{1,3,4,5\}$ is not a circuit because $\{1,4,5\}$ is not an independent set. 
\end{example}

\subsection{Designs}
A design is an ordered tuple $(X,\mathcal{A})$ where $X$ is any finite set (called the set of points) and $\mathcal{A} \subseteq 2^X$ such that all sets in $\mathcal{A}$ are of the same size. The elements of $\mathcal{A}$ are called blocks. A parallel class $\mathcal{P}$ of $(X,\mathcal{A})$ is a subset of $\mathcal{A}$ such that $\mathcal{P}$ is a partition of $X$. 
\begin{example}
Let $X = \{1,2,3,4,5,6\}$, $\mathcal{A} = \{\{1,2\}, \{1,3\}, \{2,5\}, \{3,4\}, \{3,5\}, \{4,6\}, \{5,6\}\}$. It can be seen that $(X,\mathcal{A})$ is a design. Let
\begin{IEEEeqnarray*}{l}
\mathcal{P}_1 = \{\{1,2\}, \{3,4\}, \{5,6\}\}\\
\mathcal{P}_2 = \{\{1,3\}, \{2,5\}, \{4,6\}\}\\
\mathcal{P}_3 = \{\{1,2\}, \{4,6\}, \{3,5\}\}.
\end{IEEEeqnarray*} 
Sets $\mathcal{P}_1, \mathcal{P}_2, \mathcal{P}_3$ are parallel classes of $(X,\mathcal{A})$ as each of these sets is a partition of $X$. 
\end{example}

\section{Constructing designs from matrices}\label{construct}
In \cite{rama}, Tang and Ramamoorthy showed a method to construct a combinatorial structure called resolvable designs from matrices. Our method to construct designs is similar to this method in \cite{rama}; however, we construct designs, which are less restrictive than resolvable designs (every resolvable design is a design, but a design may not be a resolvable design). 

Let $G$ be an $k \times m$ matrix over a finite field $\mathbb{F}_q$ where $G$ has no all zero row. Let $Q$ be an $m \times q^m$ matrix over $\mathbb{F}_q$ with no two identical columns. Hence, every vector in $\mathbb{F}_q^m$ is a column of $Q$. Let $D = G \times Q$. Define
\begin{IEEEeqnarray}{l}
\text{for } 1\leq i\leq k, 0\leq j\leq q-1:\, B(i,j) = \{l \,|\, D[i,l] = j \}.\label{def1}
\end{IEEEeqnarray} 
The following Lemma~\ref{lemmaRD} has been, in essence, proved by Tang and Ramamoorthy in \cite{rama}. We provide an independent proof by establishing a connection between cosets of a group homomorphism and blocks of a design. 
\begin{lemma}\label{lemmaRD}
For $X_G = \{1,2,\ldots,q^m\}$, $\mathcal{A}_G = \{B(i,j)\,|\, 1\leq i\leq k, 0\leq j\leq q-1\}$ (where $B(i,j)$ is defined in (\ref{def1})), $(X_G,\mathcal{A}_G)$ is a  design with $k$ parallel classes $\mathcal{P}_i = \{B(i,j)\,|\, 0\leq j\leq q-1 \}$ and each block $B(i,j)$ contains $q^{m-1}$ elements for $1\leq i\leq k$, $0\leq j\leq q-1$.
\end{lemma} 
The proof can be found in Appendix~\ref{appendixRD}.

\begin{example}\label{example1}
\begin{IEEEeqnarray*}{l}
G = \begin{bmatrix}
1 & 0 & 0\\
0 & 1 & 0\\
0 & 0 & 1\\
1 & 1 & 1
\end{bmatrix},\;\;
Q = \begin{bmatrix}
0 & 0 & 0 & 0 & 1 & 1 & 1 & 1\\
0 & 0 & 1 & 1 & 0 & 0 & 1 & 1\\
0 & 1 & 0 & 1 & 0 & 1 & 0 & 1
\end{bmatrix},\;\;
D = \begin{bmatrix}
0 & 0 & 0 & 0 & 1 & 1 & 1 & 1\\
0 & 0 & 1 & 1 & 0 & 0 & 1 & 1\\
0 & 1 & 0 & 1 & 0 & 1 & 0 & 1\\
0 & 1 & 1 & 0 & 1 & 0 & 0 & 1
\end{bmatrix}.
\end{IEEEeqnarray*}
Here $G$ is a $4 \times 3$ matrix over $\mathbb{F}_2$. We have:
\begin{IEEEeqnarray*}{l}
B(1,0) = \{1,2,3,4\},\; B(1,1) = \{5,6,7,8\},\; B(2,0) = \{1,2,5,6\},\; B(2,1) = \{3,4,7,8\},\\ B(3,0) = \{1,3,5,7\},\; B(3,1) =\{2,4,6,8\},\; B(4,0) = \{1,4,6,7\},\; B(4,1) =\{2,3,5,8\}.\\
\mathcal{P}_1 = \{\{1,2,3,4\},\{5,6,7,8\}\},\;
\mathcal{P}_2 = \{1,2,5,6\},\{3,4,7,8\}\},\;
\mathcal{P}_3 = \{1,3,5,7\},\{2,4,6,8\}\},\\
\mathcal{P}_4 = \{1,4,6,7\},\{2,3,5,8\}\}.
\end{IEEEeqnarray*}
Let $X_G = \{1,2,3,4,5,6,7,8\}$, $\mathcal{A}_G = \{B(1,0), B(1,1), B(2,0), B(2,1), B(3,0), B(3,1),\newline B(4,0), B(4,1)\}$. It can be seen that $(X_G, \mathcal{A}_G)$ is indeed a  design. Furthermore, each set $\mathcal{P}_i$ for $i=1,2,3,4$ partitions $X_G$, so $\mathcal{P}_i$ must be a parallel class.
\end{example}

\begin{lemma}\label{1lemma}
$D$ has $q^{rank(G)}$ distinct column vectors, each of which appears in $q^{m - rank(G)}$ distinct columns.
\end{lemma}
See Appendix~\ref{appendix1lemma} for proof. In Example~\ref{example1}, in accordance with Lemma~\ref{1lemma}, it can be seen that $D$ has $2^3$ (since $rank(G)=3$) distinct columns, each of which appears once.
\begin{corollary}\label{1coro}
If $k = m$ and $rank(G) = m$, then $B(1,j_1) \cap B(2,j_2) \cap \cdots \cap B(m,j_m)$ is a singleton set for any $0\leq j_1,j_2,\ldots,j_m\leq q-1$, and every element in $\{1,2,\ldots,q^m\}$ can be expressed as an intersection of $m$ unique blocks no two of which belong to the same parallel class.
\end{corollary}
See Appendix~\ref{appendix1coro} for proof.
%

\begin{corollary}\label{2coro}
If $k = m-1$ and $rank(G) = m-1$, then for any $0\leq j_1,j_2,\ldots,j_m \leq q-1$, $B(1,j_1) \cap B(2,j_2) \cap \cdots \cap B(m-1,j_{m-1})$ contains $q$ elements. 
\end{corollary}
See Appendix~\ref{appendix2coro} for proof.

\begin{corollary}\label{3coro}
If $k = m+1$ and $\{1,2,\ldots,k\}$ is a circuit of $G$, then for any $0\leq j_1,j_2,\ldots,j_m \leq q-1$ there exists only one block $B(m+1,j_{m+1}) \in \mathcal{P}_{m+1}$ such that $B(1,j_1) \cap B(2,j_2) \cap \cdots \cap B(m,j_m) \cap B(m+1,j_{m+1})$ is non-empty.
\end{corollary}
See Appendix~\ref{appendix3coro} for proof.

\section{Shared cache coded caching scheme}\label{scheme}
%
%
Consider an SC-CC problem with $\Lambda$ caches, $N$ files, $\frac{M}{N} = \frac{t}{q}$, and $K$ users, where $q$ is a power of a prime and $t\in \{1,2,\ldots,q\}$. Our objective is to construct an SC-CC scheme for a given constraint on the subpacketization level. Say the maximum allowed subpacketization level is $F_{\text{max}}$. 

Define $n = \ceil{\frac{\Lambda}{q}}$. Let $m$ be an integer such that $m \geq 2$, $q^m \leq F_{\text{max}}$, $m \leq n-1$. Say there exists an $n \times m$ matrix $G$ over $\mathbb{F}_q$ such that (i) $rank(G) = m$; (ii) given $\mathcal{C} = \{C_1,C_2,\ldots,C_{x}\}$ is the set of all $(m+1)$-length circuits of $G$ (where $x = |\mathcal{C}|$), every element in $\{1,2,\ldots,n\}$ belongs to at least one circuit in $\mathcal{C}$.

Let $(X_G,\mathcal{A}_G)$ be the corresponding  design constructed from $G$ using the procedure stated in Section~\ref{construct}. We know that $(X_G,\mathcal{A}_G)$ has $n$ parallel classes each having $q$ blocks. We denote the parallel classes as $\mathcal{P}_1, \mathcal{P}_2, \ldots, \mathcal{P}_n$. Let the blocks contained in $\mathcal{P}_i$ be denoted by $B(i,0), B(i,1), \ldots, B(i,q-1)$ for $1\leq i\leq n$.

Choose any $(n-1)q$ caches and arbitrarily label each of them as $c_{(i,j)}$ for $1\leq i\leq n-1, 0\leq j\leq q-1$ without using any label twice. The rest of the $\Lambda - (n-1)q$ caches are labeled arbitrarily by choosing a label from $c_{(n,0)}, c_{(n,1)}, \ldots, c_{(n,\Lambda - (n-1)q - 1)}$ without choosing any label twice. It can be seen that since $n \geq \frac{\Lambda}{q}$ implies $\Lambda - nq \leq 0$, we have $\Lambda - (n-1)q - 1\leq q - 1$. 
\subsection{Placement}\label{placement:sc-cc}
Let the $N$ files at the server be denoted by $W^i$ for $1\leq i\leq N$. Split each file $W^i$ into $q^m$ disjoint subfiles of equal size. The subfiles of the file $W^i$ are denoted as $W^i_{1}, W^i_{2}, \ldots, W^i_{q^m}$ for $1\leq i\leq N$. The subscript $j$  for $1\leq j\leq q^m$ is called the index of the subfile $W^i_j$. 

Define $Z_{(i,j)} = \{B(i,j), B(i,(j+1)_q), B(i,(j+2)_q), \ldots, B(i,(j+t-1)_q)\}$ where $0\leq j\leq q-1$, $1\leq i\leq n$, $1\leq t\leq q$. Cache $c_{(i,j)}$ (if it exists) stores a subfile (of all files) if the index of the subfile belongs to a block contained in $Z_{(i,j)}$. So each cache stores $tq^{m-1}$ subfiles of each file. Since each file consists of $q^m$ subfiles, we have $M = \frac{N t}{q}$ files, or $\frac{M}{N} = \frac{t}{q}$.

\subsection{User-to-cache association}
\begin{figure}[htpb]
\centering
\includegraphics[width=0.9\textwidth]{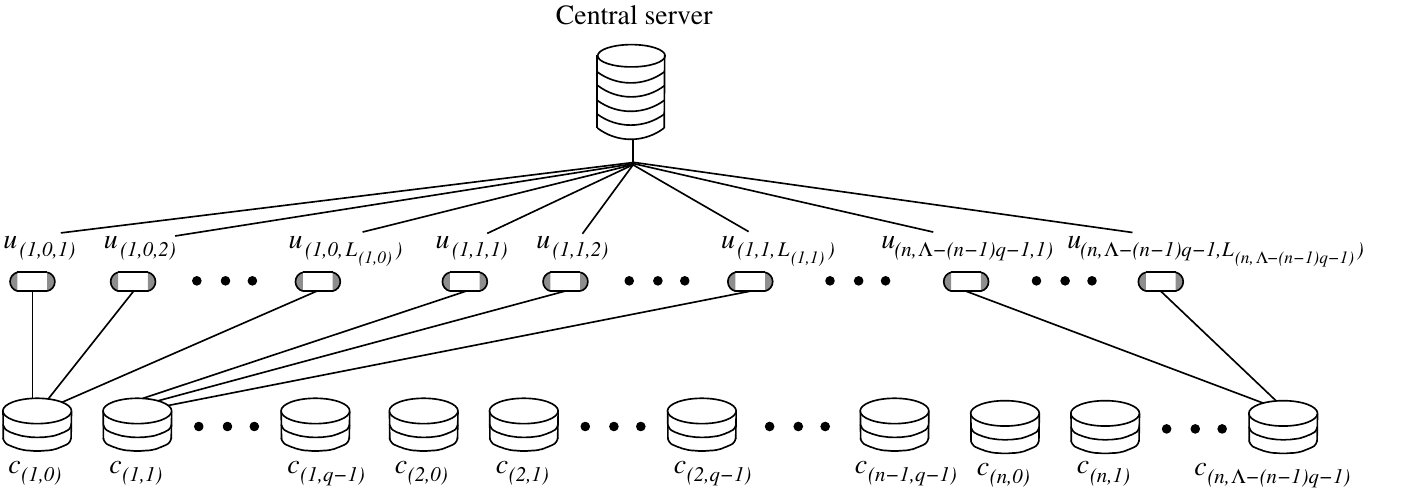}
\caption{A pictorial description of the SC-CC problem along with some of the labelings and notations considered to describe our scheme in Section~\ref{scheme}. A user $u_{(i,j,z)}$ accesses the cache $c_{(i,j)}$--the same is indicated by an edge connecting $u_{(i,j,z)}$ and $c_{(i,j)}$.}
\label{graph2}
\end{figure}

The user-to-cache association is independent of the placement. Let the set of users accessing the cache $c_{(i,j)}$ be denoted by $U_{(i,j)}$ and define $L_{(i,j)} = |U_{(i,j)}|$ (so $L_{(i,j)}$ is the number of users that accesses the cache $c_{(i,j)}$). The users in $U_{(i,j)}$ are arbitrarily labeled as $u_{(i,j,z)}$ for $1\leq z\leq L_{(i,j)}$ with no label being used twice. A figure depicting the SC-CC problem with some of the notations is shown in Fig.~\ref{graph2}.

\subsection{Delivery}\label{delivery:sc-cc}
Let the file demanded by the user $u_{(i,j,z)}$ be denoted by $W^{d(i,j,z)}$ where $d(i,j,z) \in \{1,2,\ldots,N\}$.

Say for some $1 \leq \alpha_1 <\alpha_2 < \cdots < \alpha_{m}  \leq n$ the row vectors $G[\alpha_i,:]$ for $1\leq i\leq m$ are independent over $\mathbb{F}_q$. Then, as per Corollary~\ref{1coro}, $B(\alpha_1,l_1) \cap B(\alpha_2,l_2) \cap \cdots \cap B(\alpha_m,l_m)$ contains only one element for $0\leq l_1,l_2,\ldots,l_m\leq q-1$. Define 
\begin{IEEEeqnarray*}{l}
e_{\{(\alpha_1,l_1),(\alpha_2,l_2),\ldots,(\alpha_m,l_m)\}} = B(\alpha_1,l_1) \cap B(\alpha_2,l_2) \cap \cdots \cap B(\alpha_m,l_m).
\end{IEEEeqnarray*}

Consider any circuit $C_y \in \mathcal{C}$ where $y \in \{1,2,\ldots,x\}$. Say $C_y = \{\alpha_1,\alpha_2,\ldots,\alpha_{m+1}\}$ where $1 \leq \alpha_1 <\alpha_2 < \cdots < \alpha_{m+1}  \leq n$. 

As any $m$ rows of $G[C_y,:]$ are independent, the matrix $G[C_y,:]$ is full rank. 
Since each element in $\{1,2,\ldots,q^m\}$ can be represented as an intersection of $m$ unique blocks chosen from $m$ distinct parallel classes $\mathcal{P}_{\alpha_1}, \mathcal{P}_{\alpha_2}, \ldots, \mathcal{P}_{\alpha_m}$ where one block is chosen from each parallel class, as per Corollary~\ref{1coro} we have
\begin{IEEEeqnarray*}{l}
\{1,2,\ldots,q^k \} = \{e_{\{(\alpha_1,l_1),(\alpha_2,l_2),\ldots,(\alpha_m,l_m)\}} \,|\, 0\leq l_1,l_2,\ldots,l_m\leq q-1\}
\end{IEEEeqnarray*}
Define a $q^m \times m$ matrix $A_{C_y}$ for the circuit $C_y$ as following:
\begin{IEEEeqnarray*}{l}
\text{For } 1\leq j\leq m, 0\leq l_j \leq q-1:\; A_{C_y}[e_{\{(\alpha_1,l_1),(\alpha_2,l_2),\ldots,(\alpha_m,l_m)\}},j] = l_j
\end{IEEEeqnarray*}
For $C_y$ we define $mq^{m-1}$ sets $E_{\{(\alpha_1,l_1),(\alpha_2,l_2),\ldots,(\alpha_{i-1},l_{i-1}),(\alpha_{i+1},l_{i+1}),\ldots,(\alpha_m,l_m)\}}^{C_y}$ for $1\leq i\leq m$, $0\leq l_1,l_2,\ldots,l_{i-1},l_{i+1},\ldots,l_m\leq q-1$ in the following way:
\begin{IEEEeqnarray*}{l}
E_{\{(\alpha_1,l_1),(\alpha_2,l_2),\ldots,(\alpha_{i-1},l_{i-1}),(\alpha_{i+1},l_{i+1}),\ldots,(\alpha_m,l_m)\}}^{C_y}\;\; \\\hfill =\> B(\alpha_1,l_1)\cap B(\alpha_2,l_2)\cap\cdots \cap B(\alpha_{i-1},l_{i-1})\cap B(\alpha_{i+1},l_{i+1})\cap \cdots\cap B(\alpha_m,l_m)\}.
\end{IEEEeqnarray*}
%
For each set $E_{\{(\alpha_1,l_1),(\alpha_2,l_2),\ldots,(\alpha_{i-1},l_{i-1}),(\alpha_{i+1},l_{i+1}),\ldots,(\alpha_m,l_m)\}}^{C_y}$ we further define $q$ sets \newline $E_{\{(\alpha_1,l_1),(\alpha_2,l_2),\ldots,(\alpha_{i-1},l_{i-1}),(\alpha_{i+1},l_{i+1}),\ldots,(\alpha_m,l_m)\}}^{\{C_y, (\alpha_i,l_i)\}}$ for $0\leq l_i\leq q-1$ in the following way: \newline $E_{\{(\alpha_1,l_1),(\alpha_2,l_2),\ldots,(\alpha_{i-1},l_{i-1}),(\alpha_{i+1},l_{i+1}),\ldots,(\alpha_m,l_m)\}}^{\{C_y, (\alpha_i,l_i)\}}$ contains all elements that are in \newline $E_{\{(\alpha_1,l_1),(\alpha_2,l_2),\ldots,(\alpha_{i-1},l_{i-1}),(\alpha_{i+1},l_{i+1}),\ldots,(\alpha_m,l_m)\}}^{C_y}$ but not in $Z_{(\alpha_i,l_i)}$ where $1\leq i\leq m$.
In Appendix~\ref{prop:C_y} we show several properties related to $C_y$ and the  design $(X_G,\mathcal{A}_G)$. These properties will be useful in proving  Theorem~\ref{thmx} shown later in this section. 

We want to find out the $q-t$ blocks in $\mathcal{P}_{\alpha_{m+1}}$ that contains an element from \newline $E_{\{(\alpha_1,l_1),(\alpha_2,l_2),\ldots,(\alpha_{i-1},l_{i-1}),(\alpha_{i+1},l_{i+1}),\ldots,(\alpha_m,l_m)\}}^{\{C_y, (\alpha_i,l_i)\}}$ -- we want to store the indices of all these blocks in a vector. Towards this goal, 
we construct a row vector $J_{\{(\alpha_1,l_1),(\alpha_2,l_2),\ldots,(\alpha_{i-1},l_{i-1}),(\alpha_{i+1},l_{i+1}),\ldots,(\alpha_m,l_m)\}}^{\{C_y,(\alpha_i,l_i)\}}$ for $0\leq l_1,l_2,\ldots,l_m \leq q-1$, $1\leq i\leq m$ as per \textbf{Algorithm~\ref{algox1}}.
\begin{algorithm}
\caption{Construction of $J_{\{(\alpha_1,l_1),(\alpha_2,l_2),\ldots,(\alpha_{i-1},l_{i-1}),(\alpha_{i+1},l_{i+1}),\ldots,(\alpha_m,l_m)\}}^{\{C_y,(\alpha_i,l_i)\}}$.}\label{algox1}
\begin{algorithmic}[1]
\State In the parallel class $\mathcal{P}_{\alpha_{m+1}}$ find the block to which the element $e_{\{(\alpha_1,l_1),(\alpha_2,l_2),\ldots,(\alpha_m,l_m)\}}$ belong. Say for some $0\leq l_{m+1} \leq q-1$, $e_{\{(1,l_1),(2,l_2),\ldots,(m,l_m)\}} \in B(\alpha_{m+1},l_{m+1})$.
\State $k \gets 1$, $j \gets l_{m+1}+1$.
\While{$k \leq q-t$}
\If{$|B(\alpha_{m+1},(j)_q) \cap E_{\{(\alpha_1,l_1),(\alpha_2,l_2),\ldots,(\alpha_{i-1},l_{i-1}),(\alpha_{i+1},l_{i+1}),\ldots,(\alpha_m,l_m)\}}^{\{C_y, (\alpha_i,l_i)\}}| == 1$}
\State $J_{\{(\alpha_1,l_1),(\alpha_2,l_2),\ldots,(\alpha_{i-1},l_{i-1}),(\alpha_{i+1},l_{i+1}),\ldots,(\alpha_m,l_m)\}}^{\{C_y,(\alpha_i,l_i)\}}[k] \gets (j)_q$.
\State $k \gets k+1$.
\EndIf
\State $j \gets j+1$.
\EndWhile
\end{algorithmic}
\end{algorithm}

In Appendix~\ref{prop:J} we show several properties of $J_{\{(\alpha_1,l_1),(\alpha_2,l_2),\ldots,(\alpha_{i-1},l_{i-1}),(\alpha_{i+1},l_{i+1}),\ldots,(\alpha_m,l_m)\}}^{\{C_y,(\alpha_i,l_i)\}}$. 

Construct an $n \times q$ matrix $S$ where $S[i,j+1] = L_{(i,j)}$ for $1\leq i\leq n-1, 0\leq j\leq q-1$; $S[n,j+1] = L_{(n,j)}$ for  $0\leq j\leq \Lambda - (n-1)q -1$; and $S[n,j+1] = 0$ for  $\Lambda - (n-1)q\leq j\leq q-1$. Given a set $\Omega \subseteq \{1,2,\ldots,n\}$ define $K_{\Omega} = \sum_{i \in \Omega} \sum_{j=0}^{q-1} S[i,j+1]$. 

\begin{algorithm}
\caption{Delivery algorithm.}\label{algox2}
\begin{algorithmic}[1]
\State $r \gets 0$.
\While{$S[i,j] \neq 0$ for some $1\leq i\leq n$, $1\leq j\leq q$}\label{linewhile}
\State Compute $K_{C_1}, K_{C_2}, \ldots, K_{C_x}$. Select a circuit $\bar{C} \in \{C_1,C_2,\ldots,C_{x}\}$ such that $K_{\bar{C}} \geq K_{C_l}$ \hspace*{15pt} for $1\leq l\leq x$. Say $\bar{C} = \{\beta_1,\beta_2,\ldots,\beta_{m+1}\}$ where $1 \leq \beta_1 <\beta_2 < \cdots < \beta_{m+1}  \leq n$. 
%
%
\State $sum \gets 0$, $flag \gets 0$.
\For{$a \gets 1,q^m$}\label{linex2}
\State $l_1 \gets A_{\bar{C}}[a,1]$, $l_2 \gets A_{\bar{C}}[a,2]$, $\ldots$, $l_m \gets A_{\bar{C}}[a,m]$. (note $e_{\{(\beta_1,l_1),(\beta_2,l_2),\ldots,(\beta_m,l_m)\}} = a$)\label{linex1}
\For{$j \gets 1, q-t$}\label{linexx1}
\For{$i \gets 1, m$}
\If{$S[\beta_i,l_i+1] \neq 0$} 
\State $flag \gets 1$, $l^* \gets J_{\{(\beta_1,l_1),(\beta_2,l_2),\ldots,(\beta_{i-1},l_{i-1}),(\beta_{i+1},l_{i+1}),\ldots,(\beta_m,l_m)\}}^{\{\bar{C},(\beta_i,l_i)\}}[j]$.\label{line1}
\State $sum \gets sum + W^{d(\beta_i,l_i,S[\beta_i,l_i+1])}_{e_{\{(\beta_1,l_1),(\beta_2,l_2),\ldots,(\beta_{i-1},l_{i-1}),(\beta_{i+1},l_{i+1}),\ldots,(\beta_m,l_m),(\beta_{m+1},l^*)\}}}$.
\EndIf
\EndFor
\State Find the block in $\mathcal{P}_{\beta_{m+1}}$ to which $a$ belong. Say $a \in B(\beta_{m+1},l_{m+1})$ for some \hspace*{50pt} $0\leq l_{m+1} \leq q-1$.\label{line3} 
\If{$S[\beta_{m+1},(l_{m+1} + j)_q+1] \neq 0$}
\State $flag \gets 1$, $sum \gets sum + W^{d(\beta_{m+1},(l_{m+1} + j)_q), S[\beta_{m+1},(l_{m+1} + j)_q + 1])}_{e_{\{(\beta_1,l_1),(\beta_2,l_2),\ldots,(\beta_m,l_m)\}}}$.
\EndIf
\If{$flag == 1$} \label{liney1}  
\State $r \gets r + 1$. Broadcast $sum$.\label{line2} 
\EndIf
\EndFor
\EndFor
\For{$i \gets 1, m+1$}
\For{$l \gets 0, q-1$}
\State $S[\beta_i,l+1] \gets max\{0, S[\beta_i,l+1] - 1\}$.
\EndFor
\EndFor
\EndWhile
\State Achievable rate $R = \frac{r}{q^m}$.
\end{algorithmic}
\end{algorithm}
\begin{theorem}\label{thmx}
Consider an SC-CC problem with $\Lambda$ caches, $N$ files, $\frac{M}{N} = \frac{t}{q}$, and $K$ users, where $q$ is a power of a prime and $t\in \{1,2,\ldots,q\}$. For $n = \ceil{\frac{\Lambda}{q}}$, let $G$ be an $n \times m$ matrix over $\mathbb{F}_q$ such that: (i) $2 \leq m \leq n-1$, (ii) $rank(G) = m$, (iii) $q^m \leq F_{\text{max}}$, (iv) every element in $\{1,2,\ldots,n\}$ is a member of at least one $(m+1)$-length circuit of $G$; then a rate $R = \frac{r}{q^m}$ files is achievable for a subpacketization level $q^m$ where $r$ is given by \textbf{Algorithm~\ref{algox2}}. Furthermore, new caches can be added without requiring any changes to the placement of the existing caches.
\end{theorem}
The proof of Theorem~\ref{thmx} is shown in Appendix~\ref{proof:thmx}. Since $2 \leq m \leq n-1$, we have $n-1 \geq 2$, or, $n \geq 3$. As $n = \ceil{\frac{\Lambda}{q}}$ and $q \geq 2$, we must have $\Lambda \geq 5$ for the Theorem~\ref{thmx} to be applicable.
 
It can be shown that the matrix $G$ in Theorem~\ref{thmx} always exists: there  always exists $m$ independent rows, sum these rows and one can get a summed row, this summed row and the $m$ independent rows form an $(m+1)$-length circuit; for the next $m$ rows use the same independent rows, these rows and the summed row forms another $(m+1)$-length circuit; repeat this process until less than $m$ rows remain; then choose some rows among the $m$ independent rows for the remaining rows, these rows and the rows not chosen (among the $m$ rows) and the summed row forms an $(m+1)$-length circuit. For all examples of our scheme shown in this paper, we have followed this same procedure to construct $G$.


\subsection{Adding new caches}\label{disturb}
In a practical scenario, a cache may be offline during the placement phase for various reasons. In our SC-CC scheme, new caches can be added without unsettling the placement of the existing caches. Suppose $\Delta$ caches are to be added anytime after conducting the placement. We describe how our scheme can be re-arranged accordingly.
%
\begin{case}
$(\Delta)_q \leq q - (\Lambda)_q$.
\end{case}
Define $n^\prime = \floor{\frac{\Delta}{q}}$, and let $G^\prime$ be an $n^\prime \times m$ matrix over $\mathbb{F}_q$ such that every element in $\{1,2,\ldots,n+n^\prime\}$ is a member of at least one $(m+1)$-length circuit of the matrix $[G; G^\prime]$.


Among the $\Delta$ new caches, choose any $(\Delta)_q$ caches, and arbitrarily label them as $c_{(n,j)}$ for $\Lambda - (n-1)q \leq j\leq \Lambda - (n-1)q + (\Delta)_q - 1$ without choosing any label twice. Note, we have $(\Lambda)_q = \Lambda - (n-1)q$,  so from the case condition we have $\Lambda - (n-1)q + (\Delta)_q - 1\leq q - 1$. The remaining caches in $\Delta$ are arbitrarily labeled as  $c_{(i,j)}$ for $n+1 \leq i\leq n+n^\prime$, $0\leq j\leq q-1$ without choosing any label twice.

For $\Lambda - (n-1)q \leq j\leq \Lambda - (n-1)q + (\Delta)_q - 1$, cache $c_{(n,j)}$ stores a subfile (of all files) if the index of the subfile belongs to a block contained in $Z_{(i,j)}$ which is defined in Section~\ref{placement:sc-cc}.

Let $(X_{G^\prime},\mathcal{A}_{G^\prime})$ be the corresponding  design constructed from $G^\prime$ using the procedure stated in Section~\ref{construct}. We know $(X_{G^\prime},\mathcal{A}_{G^\prime})$ has $n^\prime$ parallel classes each having $q$ blocks. From $(X_{G^\prime},\mathcal{A}_{G^\prime})$, define $Z_{(i,j)}$ for $n+1 \leq i\leq n+n^\prime$, $0\leq j\leq q-1$ analogously to Section~\ref{placement:sc-cc}. For $n+1 \leq i\leq n+n^\prime$, $0\leq j\leq q-1$, cache $c_{(i,j)}$ stores a subfile (of all files) if the index of the subfile belongs to a block contained in $Z_{(i,j)}$.

The delivery process is similar to Section~\ref{delivery:sc-cc} (replace $n$ by $n+n^\prime$ and $G$ by $[G;G^\prime]$). 
\begin{case}
$(\Delta)_q > q - (\Lambda)_q$.
\end{case}
%
Define $n^\prime = \ceil{\frac{\Delta}{q}}$, and caches in $\Delta$ are arbitrarily labeled as  $c_{(i,j)}$ for $n+1 \leq i\leq n+n^\prime$, $0\leq j\leq q-1$ without choosing any label twice. The rest is similar to \textit{Case} 1.

\setcounter{case}{0}
\subsection{Examples}\label{examples}
We consider the same SC-CC problem shown in Example 2 of references~\cite{peter} and \cite{peter2}.
\begin{example}\label{exam1}
Consider an SC-CC problem with $\Lambda = 9$ caches, $K = 45$ users, \newline $\mathcal{L} = \{8,7,6,6,5,4,4,3,2\}$, $N$ files. We study the example for $\frac{M}{N} \in \{\frac{1}{3}, \frac{2}{3}\}$. Reference~\cite{peter} studies the same example for $\frac{M}{N} = \frac{2}{3}$, and reference~\cite{peter2} studies it for $\frac{M}{N} = \frac{1}{3}$. 


We choose $q = 3$. So $n = 3$, and hence we have $m = 2$. Let $G$ be the following $3 \times 2$ matrix of rank $2$ over $\mathbb{F}_3$, shown along with respective matrices $Q$ and $D$.
\begin{IEEEeqnarray*}{l}
G = \begin{bmatrix}
1 & 0\\
0 & 1\\
1 & 1
\end{bmatrix}, 
Q = \begin{bmatrix}
0 & 0 & 0 & 1 & 1 & 1  & 2 & 2 & 2\\
0 & 1 & 2 & 0 & 1 & 2 & 0 & 1 & 2
\end{bmatrix}, 
D = \begin{bmatrix}
0 & 0 & 0 & 1 & 1 & 1 & 2 & 2 & 2\\
0 & 1 & 2 & 0 & 1 & 2 & 0 & 1 & 2\\
0 & 1 & 2 & 1 & 2 & 0 & 2 & 0 & 1
\end{bmatrix}.
\end{IEEEeqnarray*}
We obtain the design $(X_G,\mathcal{A}_G)$ with $X_G = \{1,2,\ldots,9\}$, $\mathcal{A} = \{B(i,j)\,|\, 1\leq i\leq 3, 0\leq j\leq 2\}$ where $B(i,j) = \{l \,|\, D[i,l] = j \}$. So we have $B(1,0) = \{1,2,3\}, B(1,1) = \{4,5,6\}, B(1,2) = \{7,8,9\}, B(2,0) = \{1,4,7\}, B(2,1) = \{2,5,8\}, B(2,2) = \{3,6,9\}, B(3,0) = \{1,6,8\}, B(3,1) = \{2,4,9\}, B(3,2) = \{3,5,7\}$. The parallel classes are: 
\begin{IEEEeqnarray*}{l}
\mathcal{P}_1 = \{\{1,2,3\}, \{4,5,6\}, \{7,8,9\}\},\\ \mathcal{P}_2 = \{\{1,4,7\}, \{2,5,8\}, \{3,6,9\}\},\\  \mathcal{P}_3 = \{\{1,6,8\}, \{2,4,9\}, \{3,5,7\}\}.
\end{IEEEeqnarray*}

\noindent The $9$ caches are labeled as $c_{(i,j)}$ for $1\leq i\leq 3, 0\leq j\leq 2$.

\noindent\textbf{User-to-cache association:}  Let $L_{(1,0)} = 8$, $L_{(1,1)} = 6$, $L_{(1,2)} = 4$, $L_{(2,0)} = 7$, $L_{(2,1)} = 5$, $L_{(2,2)} = 3$, $L_{(3,0)} = 2$, $L_{(3,1)} = 6$, $L_{(3,2)} = 4$.

A new notation: in line~\algref{algox2}{line2} of \textbf{Algorithm~\ref{algox2}}, let $Y_{\{(\beta_1,l_1),(\beta_2,l_2),\ldots,(\beta_m,l_m)\}}^{\{j,r\}} = sum$ for the fixed values of $\beta_1,\beta_2,\ldots,\beta_m$, $l_1,l_2,\ldots,l_m$, $j$, and $r$ considered in the corresponding loop.

\begin{case}
$\frac{M}{N} = \frac{1}{3}$.
\end{case}
\noindent\textbf{Placement:} Each file $W^i$ is split into $9$ subfiles: $W^i_1, W^i_2, \ldots, W^i_9$ for $1\leq i\leq N$. Cache $c_{(i,j)}$ stores the subfile $W^l_k$ if $k$ belongs to a block contained in $Z_{(i,j)}$ for all $1\leq l\leq N$, where (since $t = 1$) $Z_{(i,j)} = \{B(i,j)\}$ for $1\leq i\leq 3$, $0\leq j\leq 2$.

\noindent\textbf{Delivery:} The matrix $G$ has only one circuit $C = \{1,2,3\}$ of length $3$. So we have  
\begin{IEEEeqnarray*}{rl}
A_{C}\> &= \begin{bmatrix}
0 & 0 & 0\\
0 & 1 & 1\\
0 & 2 & 2\\
1 & 0 & 1\\
1 & 1 & 2\\
1 & 2 & 0\\
2 & 0 & 2\\
2 & 1 & 0\\
2 & 2 & 1
\end{bmatrix}\quad
\begin{IEEEeqnarraybox}[\IEEEeqnarraystrutmode][c]{s.s.s}
$E^{C}_{\{(1,0)\}} = \{1,2,3\}$, & $E^{C}_{\{(1,1)\}} = \{4,5,6\}$, & $E^{C}_{\{(1,2)\}} = \{7,8,9\}$,\\
$E^{C}_{\{(2,0)\}} = \{1,4,7\}$, & $E^{C}_{\{(2,1)\}} = \{2,5,8\}$, & $E^{C}_{\{(2,2)\}} = \{3,6,9\}$,\\
$E_{\{(1,0)\}}^{\{C,(2,0)\}} = \{2,3\}$, & $E_{\{(1,0)\}}^{\{C,(2,1)\}} = \{1,3\}$, & $E_{\{(1,0)\}}^{\{C,(2,2)\}} = \{1,2\}$,\\
$E_{\{(1,1)\}}^{\{C,(2,0)\}} = \{5,6\}$, & $E_{\{(1,1)\}}^{\{C,(2,1)\}} = \{4,6\}$, & $E_{\{(1,1)\}}^{\{C,(2,2)\}} = \{4,5\}$,\\
$E_{\{(1,2)\}}^{\{C,(2,0)\}} = \{8,9\}$, & $E_{\{(1,2)\}}^{\{C,(2,1)\}} = \{7,9\}$, & $E_{\{(1,2)\}}^{\{C,(2,2)\}} = \{7,8\}$,\\
$E_{\{(2,0)\}}^{\{C,(1,0)\}} = \{4,7\}$, & $E_{\{(2,0)\}}^{\{C,(1,1)\}} = \{1,7\}$, & $E_{\{(2,0)\}}^{\{C,(1,2)\}} = \{1,4\}$,\\
$E_{\{(2,1)\}}^{\{C,(1,0)\}} = \{5,8\}$, & $E_{\{(2,1)\}}^{\{C,(1,1)\}} = \{2,8\}$, & $E_{\{(2,1)\}}^{\{C,(1,2)\}} = \{2,5\}$,\\
$E_{\{(2,2)\}}^{\{C,(1,0)\}} = \{6,9\}$, & $E_{\{(2,2)\}}^{\{C,(1,1)\}} = \{3,9\}$, & $E_{\{(2,2)\}}^{\{C,(1,2)\}} = \{3,6\}$.\\
$J_{\{(1,0)\}}^{\{C,(2,0)\}} = \begin{bmatrix}1 & 2\end{bmatrix},\;$ & $J_{\{(1,0)\}}^{\{C,(2,1)\}} = \begin{bmatrix}2 & 0\end{bmatrix},\;$ & $J_{\{(1,0)\}}^{\{C,(2,2)\}} = \begin{bmatrix}0 & 1\end{bmatrix},$
\end{IEEEeqnarraybox}\\
S \>&= \begin{bmatrix}
8 & 6 & 4\\
7 & 5 & 3\\
2 & 6 & 4
\end{bmatrix}\quad
\begin{IEEEeqnarraybox}[][c]{s.s.s}
$J_{\{(1,1)\}}^{\{C,(2,0)\}} = \begin{bmatrix}2 & 0\end{bmatrix},\;$ & $J_{\{(1,1)\}}^{\{C,(2,1)\}} = \begin{bmatrix}0 & 1\end{bmatrix}, \;$ & $J_{\{(1,1)\}}^{\{C,(2,2)\}} = \begin{bmatrix}1 & 2\end{bmatrix}$,\\
$J_{\{(1,2)\}}^{\{C,(2,0)\}} = \begin{bmatrix}0 & 1\end{bmatrix}$, & $J_{\{(1,2)\}}^{\{C,(2,1)\}} = \begin{bmatrix}1 & 2\end{bmatrix}$, & $J_{\{(1,2)\}}^{\{C,(2,2)\}} = \begin{bmatrix}2 & 0\end{bmatrix}$,\\
$J_{\{(2,0)\}}^{\{C,(1,0)\}} = \begin{bmatrix}1 & 2\end{bmatrix}$, & $J_{\{(2,0)\}}^{\{C,(1,1)\}} = \begin{bmatrix}2 & 0\end{bmatrix}$, & $J_{\{(2,0)\}}^{\{C,(1,2)\}} = \begin{bmatrix}0 & 1\end{bmatrix}$,\\
$J_{\{(2,1)\}}^{\{C,(1,0)\}} = \begin{bmatrix}2 & 0\end{bmatrix}$, & $J_{\{(2,1)\}}^{\{C,(1,1)\}} = \begin{bmatrix}0 & 1\end{bmatrix}$, & $J_{\{(2,1)\}}^{\{C,(1,2)\}} = \begin{bmatrix}1 & 2\end{bmatrix}$,\\
$J_{\{(2,2)\}}^{\{C,(1,0)\}} = \begin{bmatrix}0 & 1\end{bmatrix}$, & $J_{\{(2,2)\}}^{\{C,(1,1)\}} = \begin{bmatrix}1 & 2\end{bmatrix}$, & $J_{\{(2,2)\}}^{\{C,(1,2)\}} = \begin{bmatrix}2 & 0\end{bmatrix}$.
\end{IEEEeqnarraybox}\\
\end{IEEEeqnarray*}
Let us now proceed with \textbf{Algorithm~\ref{algox2}}. We have $r = 0$ to start. We see $\bar{C} = C$, $K_C = 45$, $\beta_1 = 1$, $\beta_2 = 2$, $\beta_3 = 3$, $sum = 0$, $flag = 0$. For $a = 1$, $l_1 = 0$, $l_2 = 0$. For $j = 1$, $i = 1$, we see $S[1,1] = 8 \neq 0$, so $flag = 1$, $l^* = J_{\{(2,0)\}}^{\{C,(1,0)\}}[1] = 1$. So $sum = W^{d(1,0,8)}_{4}$. For $j = 1, i = 2$, we see $S[2,1] = 7 \neq 0$, $l^* = J_{\{(1,0)\}}^{\{C,(2,0)\}}[1] = 1$. So $sum = W^{d(1,0,8)}_{4} + W^{d(2,0,7)}_{2}$. We see that in $\mathcal{P}_3$, $a = 1$ belongs to $B(3,0)$, so $l_{m+1} = 0$. $S[3,2] = 6 \neq 0$. So $sum = W^{d(1,0,8)}_{4} + W^{d(2,0,7)}_{2} + W^{d(3,1,6)}_{1}$. Since $flag = 1$,  we have $r = 1$ and $Y_{\{(1,0),(2,0)\}}^{\{1,1\}} = sum$. 

For the next iteration of the For loop in line~\algref{algox2}{linexx1}, we have $j = 2$. For $i = 1$, we see $S[1,1] = 8 \neq 0$, so $flag = 1$, $l^* = J_{\{(2,0)\}}^{\{C,(1,0)\}}[2] = 2$. So $sum = W^{d(1,0,8)}_{7}$. For $j = 2, i = 2$, we see $S[2,1] = 7 \neq 0$, $l^* = J_{\{(1,0)\}}^{\{C,(2,0)\}}[2] = 2$. So $sum = W^{d(1,0,8)}_{7} + W^{d(2,0,7)}_{3}$. In $\mathcal{P}_3$, $a = 1$ belongs to $B(3,0)$, so $l_{m+1} = 0$. $S[3,3] = 4 \neq 0$. So $sum = W^{d(1,0,8)}_{7} + W^{d(2,0,7)}_{3} + W^{d(3,2,4)}_{1}$. Since $flag = 1$, we have $r = 2$ and $Y_{\{(1,0),(2,0)\}}^{\{2,2\}} = sum$. 

Since the For loop in line~\algref{algox2}{linexx1} is over, we now have $a= 2$. We have $l_1 = 0$, $l_2 = 1$. For $j = 1$, $i = 1$, we see $S[1,1] = 8 \neq 0$, so $flag = 1$, $l^* = J_{\{(2,1)\}}^{\{C,(1,0)\}}[1] = 2$. So $sum = W^{d(1,0,8)}_{5}$. For $j = 1, i = 2$, we see $S[2,2] = 5 \neq 0$, $l^* = J_{\{(1,0)\}}^{\{C,(2,1)\}}[1] = 2$. So $sum = W^{d(1,0,8)}_{5} + W^{d(2,1,5)}_{3}$. In $\mathcal{P}_3$, $a = 2$ belongs to $B(3,1)$, so $l_{m+1} = 1$. $S[3,3] = 4 \neq 0$. So $sum = W^{d(1,0,8)}_{5} + W^{d(2,1,5)}_{3} + W^{d(3,2,4)}_{2}$. Since $flag = 1$, we have $r = 3$ and $Y_{\{(1,0),(2,1)\}}^{\{1,3\}} = sum$. Proceeding similarly, after the first iteration of the While loop in line~\algref{algox2}{linewhile} we get:
\begin{IEEEeqnarray*}{ll}
\text{For } a=1, j=1:& Y_{\{(1,0),(2,0)\}}^{\{1,1\}} = W^{d(1,0,8)}_{4} + W^{d(2,0,7)}_{2} + W^{d(3,1,6)}_{1}\\
\text{For } a=1, j=2:& Y_{\{(1,0),(2,0)\}}^{\{2,2\}} = W^{d(1,0,8)}_{7} + W^{d(2,0,7)}_{3} + W^{d(3,2,4)}_{1}\\
\text{For } a=2, j=1:& Y_{\{(1,0),(2,1)\}}^{\{1,3\}} = W^{d(1,0,8)}_{5} + W^{d(2,1,5)}_{3} + W^{d(3,2,4)}_{2}\\
\text{For } a=2, j=2:\quad & Y_{\{(1,0),(2,1)\}}^{\{2,4\}} = W^{d(1,0,8)}_{8} + W^{d(2,1,5)}_{1} + W^{d(3,0,2)}_{2}\\
\text{For } a=3, j=1:& Y_{\{(1,0),(2,2)\}}^{\{1,5\}} = W^{d(1,0,8)}_{6} + W^{d(2,2,3)}_{1} + W^{d(3,0,2)}_{3}\\
\text{For } a=3, j=2:& Y_{\{(1,0),(2,2)\}}^{\{2,6\}} = W^{d(1,0,8)}_{9} + W^{d(2,2,3)}_{2} + W^{d(3,1,6)}_{3}\\
\text{For } a=4, j=1:& Y_{\{(1,1),(2,0)\}}^{\{1,7\}} = W^{d(1,1,6)}_{7} + W^{d(2,0,7)}_{5} + W^{d(3,2,4)}_{4}\\
\text{For } a=4, j=2:& Y_{\{(1,1),(2,0)\}}^{\{2,8\}} = W^{d(1,1,6)}_{1} + W^{d(2,0,7)}_{6} + W^{d(3,0,2)}_{4}\\
\text{For } a=5, j=1:& Y_{\{(1,1),(2,1)\}}^{\{1,9\}} = W^{d(1,1,6)}_{8} + W^{d(2,1,5)}_{1} + W^{d(3,0,2)}_{5}\\
\text{For } a=5, j=2:& Y_{\{(1,1),(2,1)\}}^{\{2,10\}} = W^{d(1,1,6)}_{2} + W^{d(2,1,5)}_{2} + W^{d(3,1,6)}_{5}\\
\text{For } a=6, j=1:& Y_{\{(1,1),(2,2)\}}^{\{1,11\}} = W^{d(1,1,6)}_{9} + W^{d(2,2,3)}_{4} + W^{d(3,1,6)}_{6}\\
\text{For } a=6, j=2:& Y_{\{(1,1),(2,2)\}}^{\{2,12\}} = W^{d(1,1,6)}_{3} + W^{d(2,2,3)}_{5} + W^{d(3,2,4)}_{6}\\
\text{For } a=7, j=1:& Y_{\{(1,2),(2,0)\}}^{\{1,13\}} = W^{d(1,2,4)}_{1} + W^{d(2,0,7)}_{8} + W^{d(3,0,2)}_{7}\\
\text{For } a=7, j=2:& Y_{\{(1,2),(2,0)\}}^{\{2,14\}} = W^{d(1,2,4)}_{4} + W^{d(2,0,7)}_{9} + W^{d(3,1,6)}_{7}\\
\text{For } a=8, j=1:& Y_{\{(1,2),(2,1)\}}^{\{1,15\}} = W^{d(1,2,4)}_{2} + W^{d(2,1,5)}_{9} + W^{d(3,1,6)}_{8}\\
\text{For } a=8, j=2:& Y_{\{(1,2),(2,1)\}}^{\{2,16\}} = W^{d(1,2,4)}_{5} + W^{d(2,1,5)}_{7} + W^{d(3,2,4)}_{8}\\
\text{For } a=9, j=1:& Y_{\{(1,2),(2,2)\}}^{\{1,17\}} = W^{d(1,2,4)}_{3} + W^{d(2,2,3)}_{7} + W^{d(3,2,4)}_{9}\\
\text{For } a=9, j=2:& Y_{\{(1,2),(2,2)\}}^{\{2,18\}} = W^{d(1,2,4)}_{6} + W^{d(2,2,3)}_{8} + W^{d(3,0,2)}_{9}
\end{IEEEeqnarray*}
The For loop in line~\algref{algox2}{linex2} ends here. It can be seen that the demands of the  users $u_{(i,j,z)}$ for $1\leq i\leq 3$, $0\leq j\leq 2$, $z = L_{(i,j)}$ have been met. The matrix $S$ will now be updated by deducting $1$ from every value in $S$. Next, demands of users  $u_{(i,j,z)}$ for $1\leq i\leq 3$, $0\leq j\leq 2$, $z = L_{(i,j)} - 1$ will be met. 

Below we show how the $S$ matrix transitions with $r$; the arrow indicates the status of $S$ after a certain value of $r$. 
\begin{IEEEeqnarray*}{l}
S = \begin{bmatrix}
8 & 6 & 4\\
7 & 5 & 3\\
2 & 6 & 4
\end{bmatrix} \xrightarrow{r = 18} 
\begin{bmatrix}
7 & 5 & 3\\
6 & 4 & 2\\
1 & 5 & 3
\end{bmatrix} \xrightarrow{r = 72}
\begin{bmatrix}
4 & 2 & 0\\
3 & 1 & 0\\
0 & 2 & 0
\end{bmatrix} \xrightarrow{r = 88}
\begin{bmatrix}
3 & 1 & 0\\
2 & 0 & 0\\
0 & 1 & 0
\end{bmatrix} \xrightarrow{r = 103}
\begin{bmatrix}
2 & 0 & 0\\
1 & 0 & 0\\
0 & 0 & 0
\end{bmatrix} \xrightarrow{r = 113}\\\hfill
\begin{bmatrix}
1 & 0 & 0\\
0 & 0 & 0\\
0 & 0 & 0
\end{bmatrix}\xrightarrow{r = 119}
\begin{bmatrix}
0 & 0 & 0\\
0 & 0 & 0\\
0 & 0 & 0
\end{bmatrix}.
\end{IEEEeqnarray*}  
The transition from $r = 88$ to $r =  103$ is shown as example. After $r = 88$ we have $S[1,1] = 3, S[1,2] = 1, S[1,3] = 0, S[2,1] = 2, S[2,2] = 0, S[2,3] = 0, S[3,1] = 0, S[3,2] = 1, S[3,3] = 0$.
\begin{IEEEeqnarray*}{ll}
\text{For } a=1, j=1:& Y_{\{(1,0),(2,0)\}}^{\{1,89\}} = W^{d(1,0,3)}_{4} + W^{d(2,0,2)}_{2} + W^{d(3,1,1)}_{1}\\
\text{For } a=1, j=2:& Y_{\{(1,0),(2,0)\}}^{\{2,90\}} = W^{d(1,0,3)}_{7} + W^{d(2,0,2)}_{3} \\
\text{For } a=2, j=1:& Y_{\{(1,0),(2,1)\}}^{\{1,91\}} = W^{d(1,0,3)}_{5} \\
\text{For } a=2, j=2:\quad & Y_{\{(1,0),(2,1)\}}^{\{2,92\}} = W^{d(1,0,3)}_{8} \\
\text{For } a=3, j=1:& Y_{\{(1,0),(2,2)\}}^{\{1,93\}} = W^{d(1,0,3)}_{6} \\
\text{For } a=3, j=2:& Y_{\{(1,0),(2,2)\}}^{\{2,94\}} = W^{d(1,0,3)}_{9} +  W^{d(3,1,1)}_{3}\\
\text{For } a=4, j=1:& Y_{\{(1,1),(2,0)\}}^{\{1,95\}} = W^{d(1,1,1)}_{7} + W^{d(2,0,2)}_{5} \\
\text{For } a=4, j=2:& Y_{\{(1,1),(2,0)\}}^{\{2,96\}} = W^{d(1,1,1)}_{1} + W^{d(2,0,2)}_{6} \\
\text{For } a=5, j=1:& Y_{\{(1,1),(2,1)\}}^{\{1,97\}} = W^{d(1,1,1)}_{8} \\
\text{For } a=5, j=2:& Y_{\{(1,1),(2,1)\}}^{\{2,98\}} = W^{d(1,1,1)}_{2} +  W^{d(3,1,1)}_{5}\\
\text{For } a=6, j=1:& Y_{\{(1,1),(2,2)\}}^{\{1,99\}} = W^{d(1,1,1)}_{9} +  W^{d(3,1,1)}_{6}\\
\text{For } a=6, j=2:& Y_{\{(1,1),(2,2)\}}^{\{2,100\}} = W^{d(1,1,1)}_{3} \\
\text{For } a=7, j=1:& Y_{\{(1,2),(2,0)\}}^{\{1,101\}} =  W^{d(2,0,2)}_{8} \\
\text{For } a=7, j=2:& Y_{\{(1,2),(2,0)\}}^{\{2,102\}} =  W^{d(2,0,2)}_{9} + W^{d(3,1,1)}_{7}\\
\text{For } a=8, j=1:& Y_{\{(1,2),(2,1)\}}^{\{1,103\}} =  W^{d(3,1,1)}_{8}
%
\end{IEEEeqnarray*}
The transition from $r = 103$ to $r =  113$ is shown below. After $r = 103$ we have $S[1,1] = 2, S[1,2] = 0, S[1,3] = 0, S[2,1] = 1, S[2,2] = 0, S[2,3] = 0, S[3,1] = 0, S[3,2] = 0, S[3,3] = 0$.
\begin{IEEEeqnarray*}{ll}
\text{For } a=1, j=1:& Y_{\{(1,0),(2,0)\}}^{\{1,104\}} = W^{d(1,0,2)}_{4} + W^{d(2,0,1)}_{2}\\
\text{For } a=1, j=2:& Y_{\{(1,0),(2,0)\}}^{\{2,105\}} = W^{d(1,0,2)}_{7} + W^{d(2,0,1)}_{3} \\
\text{For } a=2, j=1:& Y_{\{(1,0),(2,1)\}}^{\{1,106\}} = W^{d(1,0,2)}_{5} \\
\text{For } a=2, j=2:\quad & Y_{\{(1,0),(2,1)\}}^{\{2,107\}} = W^{d(1,0,2)}_{8} \\
\text{For } a=3, j=1:& Y_{\{(1,0),(2,2)\}}^{\{1,108\}} = W^{d(1,0,2)}_{6} \\
\text{For } a=3, j=2:& Y_{\{(1,0),(2,2)\}}^{\{2,109\}} = W^{d(1,0,2)}_{9}\\
\text{For } a=4, j=1:& Y_{\{(1,1),(2,0)\}}^{\{1,110\}} = W^{d(2,0,1)}_{5} \\
\text{For } a=4, j=2:& Y_{\{(1,1),(2,0)\}}^{\{2,111\}} = W^{d(2,0,1)}_{6} \\
%
%
%
\text{For } a=7, j=1:& Y_{\{(1,2),(2,0)\}}^{\{1,112\}} =  W^{d(2,0,1)}_{8} \\
\text{For } a=7, j=2:& Y_{\{(1,2),(2,0)\}}^{\{2,113\}} =  W^{d(2,0,1)}_{9}
%
%
\end{IEEEeqnarray*}
Proceeding as \textbf{Algorithm~\ref{algox2}} we see that a rate $R = \frac{119}{9} \approx 13.22$ is achievable. 

Furthermore, we have observed that the achievable rate depends upon the user-to-cache association profile. For example, if the considered user-to-cache association profile had it been $L_{(3,0)} = 6$ and $L_{(3,1)} = 2$, the achievable rate would have been $\frac{120}{9}$. After trying all possible user-to-cache association profiles acceptable for this example, we found that at least $R = 14$ is always achievable. For example, for the user-to-cache association profile $L_{(1,0)} = 8$, $L_{(1,1)} = 7$, $L_{(1,2)} = 6$, $L_{(2,0)} = 6$, $L_{(2,1)} = 5$, $L_{(2,2)} = 4$, $L_{(3,0)} = 4$, $L_{(3,1)} = 3$, $L_{(3,2)} = 2$, the achievable rate is $14$.

\begin{case}
$\frac{M}{N} = \frac{2}{3}$.
\end{case}
\noindent\textbf{Placement:} Each file $W^i$ is split into $9$ subfiles: $W^i_1, W^i_2, \ldots, W^i_9$ for $1\leq i\leq N$. Cache $c_{(i,j)}$ stores the subfile $W^l_k$ if $k$ belongs to a block contained in $Z_{(i,j)}$ for any $1\leq l\leq N$, where (since $t = 2$) $Z_{(i,j)} = \{B(i,j), B(i,(j+1)_3\}$ for $1\leq i\leq 3$, $0\leq j\leq 2$.

\noindent\textbf{Delivery:} To show the delivery, we show the transition of the $S$ matrix with $r$.
\begin{IEEEeqnarray*}{l}
\begin{bmatrix}
8 & 6 & 4\\
7 & 5 & 3\\
2 & 6 & 4
\end{bmatrix} {\xrightarrow{r = 36}}
\begin{bmatrix}
4 & 2 & 0\\
3 & 1 & 0\\
0 & 2 & 0
\end{bmatrix} \xrightarrow{r = 44}
\begin{bmatrix}
3 & 1 & 0\\
2 & 0 & 0\\
0 & 1 & 0
\end{bmatrix} \xrightarrow{r = 52}
\begin{bmatrix}
2 & 0 & 0\\
1 & 0 & 0\\
0 & 0 & 0
\end{bmatrix} \xrightarrow{r = 57}
\begin{bmatrix}
1 & 0 & 0\\
0 & 0 & 0\\
0 & 0 & 0
\end{bmatrix}{\xrightarrow{r = 60}}
\begin{bmatrix}
0 & 0 & 0\\
0 & 0 & 0\\
0 & 0 & 0
\end{bmatrix}.
\end{IEEEeqnarray*} 
We see that a rate $R = \frac{60}{9} \approx 6.66$ is achievable. 

Interestingly, had it been $L_{(3,0)} = 6$ and $L_{(3,1)} = 2$, we would have the following transitions of the $S$ matrix with the values of $r$ (finally achieving  $R = \frac{59}{9} \approx 6.55$).
\begin{IEEEeqnarray*}{l}
\begin{bmatrix}
8 & 6 & 4\\
7 & 5 & 3\\
6 & 2 & 4
\end{bmatrix} {\xrightarrow{r = 36}}
\begin{bmatrix}
4 & 2 & 0\\
3 & 1 & 0\\
2 & 0 & 0
\end{bmatrix} \xrightarrow{r = 44}
\begin{bmatrix}
3 & 1 & 0\\
2 & 0 & 0\\
1 & 0 & 0
\end{bmatrix} \xrightarrow{r = 51}
\begin{bmatrix}
2 & 0 & 0\\
1 & 0 & 0\\
0 & 0 & 0
\end{bmatrix} \xrightarrow{r = 56}
\begin{bmatrix}
1 & 0 & 0\\
0 & 0 & 0\\
0 & 0 & 0
\end{bmatrix}{\xrightarrow{r = 59}}
\begin{bmatrix}
0 & 0 & 0\\
0 & 0 & 0\\
0 & 0 & 0
\end{bmatrix}.
\end{IEEEeqnarray*} 
\end{example}
\subsection{Comparison with existing works}\label{compare}
For our problem setup, the single antenna versions of the SC-CC schemes shown in references \cite{elia} and \cite{peter} are the relevant works. A general comparison of the three schemes is not possible because there is no closed form expression of the achievable rate of our scheme and the scheme in \cite{peter}. Notably, for the scheme in \cite{elia}, the subpacketization level is $\binom{\Lambda}{\Lambda\gamma}$ where $\gamma = \frac{M}{N}$. Given $\gamma = \frac{t}{q}$ where $q$ is a power prime, for any integer $m$ where $q^2 \leq q^m \leq q^{\ceil{\frac{\Lambda}{q}}-1}$, our scheme provides a solution with a subpacketization level $q^m$. So our scheme is more robust in terms of the subpacketization level. 

For the SC-CC problem considered in Example~\ref{exam1} we show a comparison of the schemes in Tables~\ref{table1z}~and~\ref{table2z}. The SC-CC problem has $\Lambda = 9$ caches, $K = 45$ users, $\mathcal{L} = \{8,7,6,6,5,4,4,3,2\}$, and $N$ files; Table~\ref{table1z} shows the comparison for $\frac{M}{N} = \frac{1}{3}$, Table~\ref{table2z} shows the comparison for $\frac{M}{N} = \frac{2}{3}$. This example has been chosen to make the comparison unbiased, as the same example has been considered in \cite{peter} and \cite{peter2} (it is the only example in \cite{peter} and \cite{peter2} where our scheme is applicable). As it has been shown in Section~\ref{examples}, for $\frac{M}{N} = \frac{1}{3}$ our scheme achieves a rate $\frac{119}{9}$ files which is strictly lesser than $14$ files achieved by the scheme in \cite{peter}, all for the same subpacketization level $9$. 
For $\frac{M}{N} = \frac{2}{3}$ our scheme has a higher rate, but has a lower subpacketization level, in comparison to the schemes in \cite{peter} and \cite{elia}. 
\begin{table}[!htbp]
\caption{Comparison with \cite{elia} and \cite{peter}. Case: $\frac{M}{N} = \frac{1}{3}$}
\label{table1z}
\centering
\begin{tabularx}{0.6\textwidth}{ | >{\centering}p{4.5cm}<{ }| >{\centering\arraybackslash$}X<{$} | >{\centering\arraybackslash$}X<{$} | }
\hline
SC-CC scheme & R & \text{Subpacketization}\\\hline
Scheme in \cite{elia} & \frac{897}{84}\approx 10.68 & 84\\\hline
Scheme in \cite{peter} (example in \cite{peter2}) & 14 & 9\\\hline

Our scheme & \frac{119}{9} \approx 13.22 & 9\\\hline
\end{tabularx}
\end{table}
\begin{table}[!htbp]
\caption{Comparison with \cite{elia} and \cite{peter}. Case: $\frac{M}{N} = \frac{2}{3}$}
\label{table2z}
\centering
\begin{tabularx}{0.5\textwidth}{ | >{\centering}p{2.5cm}<{ }| >{\centering\arraybackslash$}X<{$} | >{\centering\arraybackslash$}X<{$} | }
\hline
SC-CC scheme & R & \text{Subpacketization}\\\hline
Scheme in \cite{elia} & \frac{279}{84}\approx 3.32 & 84\\\hline
Scheme in \cite{peter} & \frac{69}{18} \approx 3.83 & 18\\\hline

Our scheme & \frac{60}{9} \approx 6.66 & 9\\\hline
\end{tabularx}
\end{table}

The SC-CC scheme in \cite{peter} shows how an existing PDA (originally intended for a DC-CC problem) can be used for an SC-CC problem. The performance of the scheme in \cite{peter} depends upon the considered PDA; we cannot confirm whether constructing the scheme in \cite{peter} using any other PDA would not yield better results.
\begin{table}[!htbp]
\caption{A $(9,9,6,9)$ PDA along with cache labels and subfile indices}
\label{tabla}
\centering
\begin{tabularx}{0.6\textwidth}{  >{\centering}X<{}| >{\centering\arraybackslash$}X<{$} | >{\centering\arraybackslash$}X<{$} | >{\centering\arraybackslash$}X<{$} |>{\centering\arraybackslash$}X<{$} |>{\centering\arraybackslash$}X<{$} |>{\centering\arraybackslash$}X<{$} |>{\centering\arraybackslash$}X<{$} |>{\centering\arraybackslash$}X<{$} |>{\centering\arraybackslash$}X<{$} |}
 & c_{(1,0)} & c_{(1,1)} & c_{(1,2)} & c_{(2,0)} & c_{(2,1)} & c_{(2,2)} & c_{(3,0)} & c_{(3,1)} & c_{(3,2)}\\\hline
$1$ & 0 & \star & \star & 1 & \star & \star & 2 & \star & \star\\\hline
$2$ & \star & 0 & \star & \star & 1 & \star & \star & 2 & \star\\\hline
$3$ & \star & \star & 0 & \star & \star & 1 & \star & \star & 2\\\hline
$4$ & 3 & \star & \star & 4 & \star & \star & 5 & \star & \star\\\hline
$5$ & \star & 3 & \star & \star & 4 & \star & \star & 5 & \star\\\hline
$6$ & \star & \star & 3 & \star & \star & 4 & \star & \star & 5\\\hline
$7$ & 6 & \star & \star & 7 & \star & \star & 8 & \star & \star\\\hline
$8$ & \star & 6 & \star & \star & 7 & \star & \star & 8 & \star\\\hline
$9$ & \star & \star & 6 & \star & \star & 7 & \star & \star & 8\\\hline
\end{tabularx}
\end{table}

\begin{figure}
\centering
\includegraphics[width=0.7\textwidth]{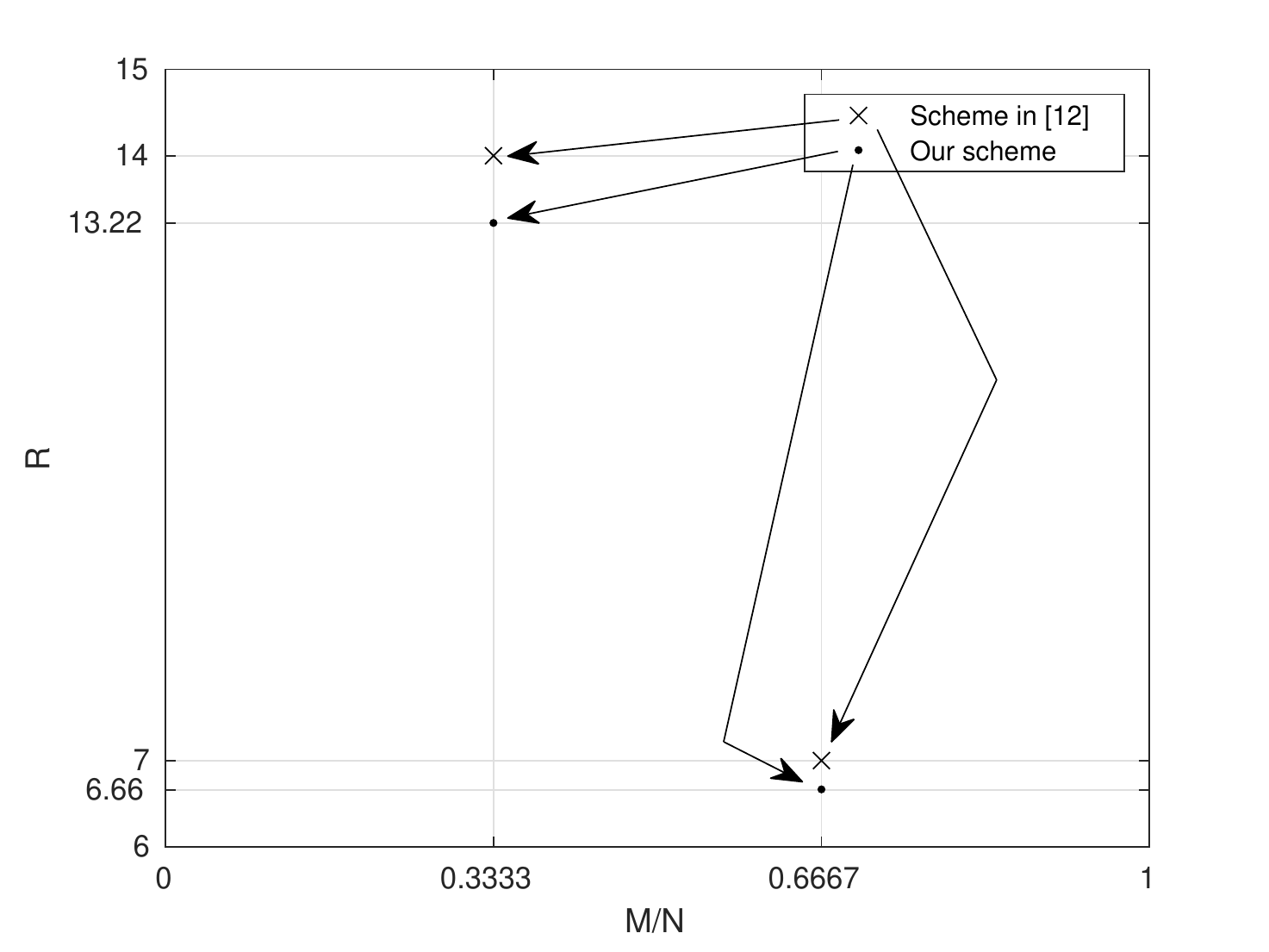}
\caption{For the SC-CC problem with $\Lambda = 9$, $K = 45$, $\mathcal{L} = \{8,7,6,6,5,4,4,3,2\}$, $F_{\text{max}} = 9$, when $\frac{M}{N} = \frac{1}{3}$, our scheme achieves a rate $\approx 13.22$ files, whereas the  scheme in \cite{peter} achieves a rate $14$ files; when $\frac{M}{N} = \frac{2}{3}$ our scheme achieves a rate $\approx 6.66$ files, whereas the  scheme in \cite{peter} achieves a rate $7$ files. This example shows that for the same subpacketization level, our scheme achieves a lesser rate in some cases.}
\label{fig:first}
\end{figure}

In Table~\ref{table2z}, the subpacketization levels of the other schemes are higher. For a better comparison, we construct the scheme in \cite{peter} using a PDA with linear subpacketization. We use a $(9,9,6,9)$ PDA recently shown by Aravind, Sarvepalli, and Thangaraj in \cite{aravind}. In Table~\ref{tabla} the PDA is reproduced from [42, Example IV.1., Table II]. Each column is labeled by a cache, and the stars in the corresponding column decides the placement of the cache; for example, for $i=1,2,3$ cache $c_{(i,0)}$ stores the subfiles indexed by $2,3,5,6,8,9$. We consider the same user-to-cache assignment as in Example~\ref{exam1}, that is, $L_{(1,0)} = 8$, $L_{(1,1)} = 6$, $L_{(1,2)} = 4$, $L_{(2,0)} = 7$, $L_{(2,1)} = 5$, $L_{(2,2)} = 3$, $L_{(3,0)} = 2$, $L_{(3,1)} = 6$, $L_{(3,2)} = 4$. Upon constructing the scheme in \cite{peter} using the PDA in Table~\ref{tabla}, we achieved a rate $R = 7$ files for a subpacketization level $9$. A comparison of our scheme and the scheme in \cite{peter} pertaining to this example is shown in Fig.~\ref{fig:first}.

Interestingly, for a different user-to-cache association, that is, $L_{(1,0)} = 8$, $L_{(1,1)} = 7$, $L_{(1,2)} = 6$, $L_{(2,0)} = 6$, $L_{(2,1)} = 5$, $L_{(2,2)} = 4$, $L_{(3,0)} = 4$, $L_{(3,1)} = 4$, $L_{(3,2)} = 2$, one could achieve a rate $R = 6$ files for  a subpacketization level $9$ using the scheme in \cite{peter} and PDA in Table~\ref{tabla}; whereas, for the same setup, our scheme achieved $R = 7$ files for a subpacketization level $9$. 

For the SC-CC problem considered in Example~\ref{exam1} we show a comparison of the schemes in Tables~\ref{table1z}~and~\ref{table2z} and in Fig.~\ref{fig:first}.

\begin{figure}
\centering
\includegraphics[width=0.7\textwidth]{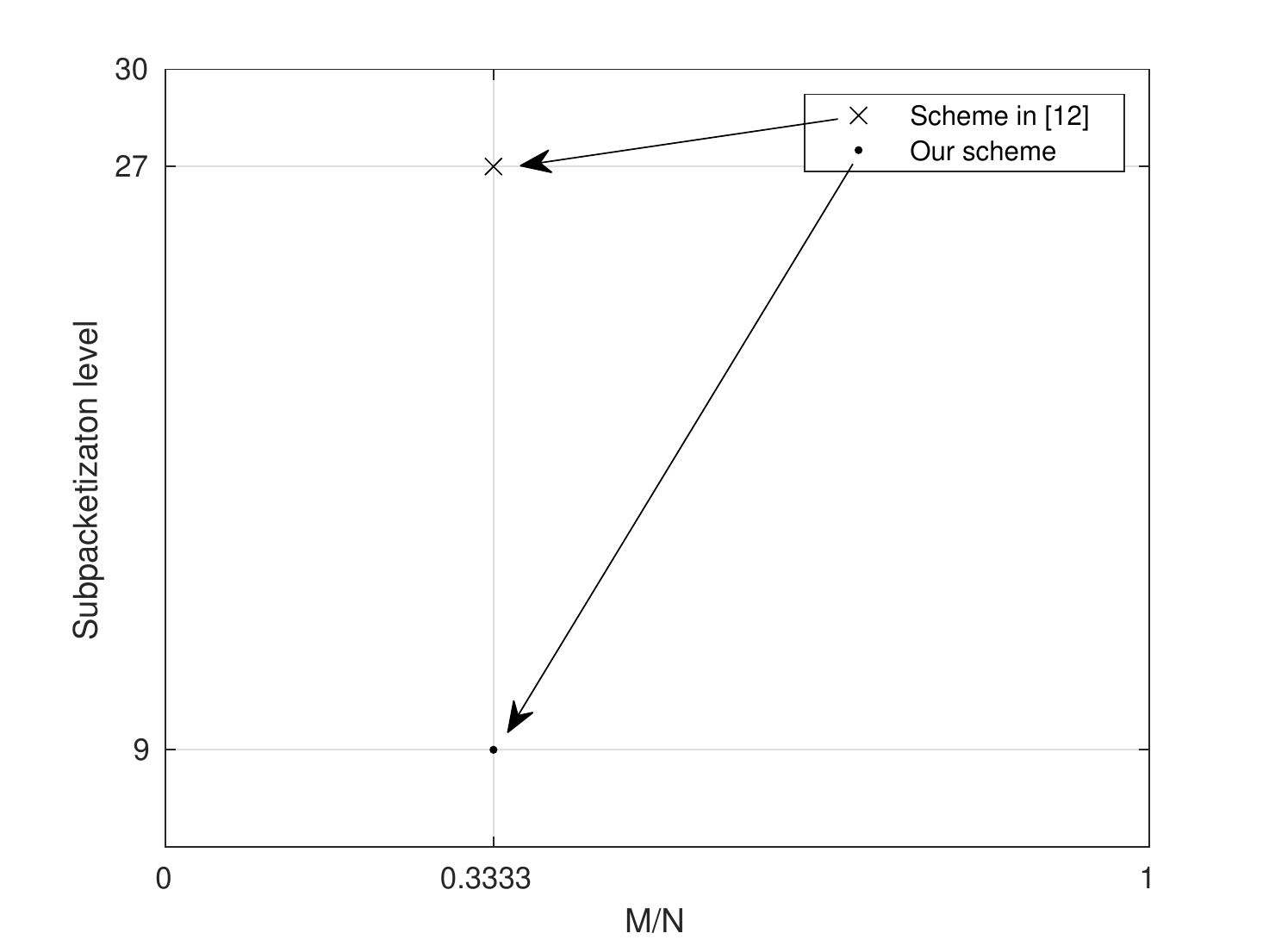}
\caption{For the SC-CC problem with $\Lambda = 12$, $K = 18$, $\mathcal{L} = \{1,1,1,2,2,2,2,2,2,1,1,1\}$, when $\frac{M}{N} = \frac{1}{3}$, our scheme achieves a rate $4$ files for a subpacketization level of $9$. In contrast, the  scheme in \cite{peter} achieves a rate $4$ files for a subpacketization level of $27$. This example shows that in some cases, our scheme uses a lesser subpacketization level to achieve the same rate as achieved by the scheme in \cite{peter}.}
\label{fig:second}
\end{figure}

We now show an example where our scheme achieves the same rate as the scheme in \cite{peter}, but using a lesser subpcketization level. Consider an SC-CC problem with $\Lambda = 12$ caches, $K = 18$ users, $\mathcal{L} = \{1,1,1,2,2,2,2,2,2,1,1,1\}$, $N$ files, $\frac{M}{N} = \frac{1}{3}$. For this problem, we show that both our scheme and the scheme in \cite{peter} achieves a rate of $4$ files, but our scheme has a subpacketization level of $9$; in contrast, the latter scheme (obtained using the PDA $\mathbf{A}^{3,3}$ shown by Yan \textit{et al.} in \cite{yctc}) has a subpacketization level of $27$. For this example problem, a comparison between our scheme and the scheme in \cite{peter} is shown in Fig.~\ref{fig:second}.
The PDA $\mathbf{A}^{3,3}$ (\textbf{Construction A} in \cite{yctc}; it is a $(12,27,9,54)$ PDA) is shown in (\ref{aq1}). It can be verified that using this PDA, the scheme in \cite{peter} achieves a rate of $4$ files.

To use our scheme for the same problem, we choose $q = 3$, $n = 4$, $m = 2$, and let $G$ be the following $4 \times 2$ matrix of rank $2$ over $\mathbb{F}_3$. The parallel classes of the design constructed from $G$ are also shown along with.
\begin{IEEEeqnarray*}{l}
G = \begin{bmatrix}
1 & 0\\
0 & 1\\
1 & 1\\
1 & 0
\end{bmatrix} \qquad
\begin{IEEEeqnarraybox}[\IEEEeqnarraystrutmode][c]{s}
$\mathcal{P}_1 = \{\{1,2,3\}, \{4,5,6\}, \{7,8,9\}\}$,\\
$\mathcal{P}_2 = \{\{1,4,7\}, \{2,5,8\}, \{3,6,9\}\}$,\\
$\mathcal{P}_3 = \{\{1,6,8\}, \{2,4,9\}, \{3,5,7\}\}$,\\
$\mathcal{P}_4 = \{\{1,2,3\}, \{4,5,6\}, \{7,8,9\}\}$.
\end{IEEEeqnarraybox}
\end{IEEEeqnarray*}
The $12$ caches are labeled as $c_{(i,j)}$ for $1\leq i\leq 4, 0\leq j\leq 3$. Let $L_{(1,0)} = L_{(1,1)} = L_{(1,2)} = 1$, $L_{(2,0)} = L_{(2,1)} = L_{(2,2)} = L_{(3,0)} = L_{(3,1)} = L_{(3,2)} = 2$, $L_{(4,0)} = L_{(4,1)} = L_{(4,2)} = 1$.
\begin{figure}
\begin{IEEEeqnarray}{l}
\mathbf{A}^{3,3} = \begin{bmatrix}
\star & 28 & 2 & \star & 30 & 6 & \star & 36 & 18 & \star & 0 & 27\\
0 & \star & 29 & \star & 31 & 7 & \star & 37 & 19 & 28 & \star & 1\\
27 & 1 & \star & \star & 32 & 8 & \star & 38 & 20 & 2 & 29 & \star\\
\star & 31 & 5 & 0 & \star & 33 & \star & 39 & 21 & 30 & \star & 3\\
3 & \star & 32 & 1 & \star & 34 & \star & 40 & 22 & 4 & 31 & \star\\
30 & 4 & \star & 2 & \star & 35 & \star & 41 & 23 & \star & 5 & 32\\
\star & 34 & 8 & 27 & 3 & \star & \star & 42 & 24 & 6 & 33 & \star\\
6 & \star & 35 & 28 & 4 & \star & \star & 43 & 25 & \star & 7 & 34\\
33 & 7 & \star & 29 & 5 & \star & \star & 44 & 26 & 35 & \star & 8\\
\star & 37 & 11 & \star & 39 & 15 & 0 & \star & 45 & 36 & \star & 9\\
9 & \star & 38 & \star & 40 & 16 & 1 & \star & 46 & 10 & 37 & \star\\
36 & 10 & \star & \star & 41 & 17 & 2 & \star & 47 & \star & 11 & 38\\
\star & 40 & 14 & 9 & \star & 42 & 3 & \star & 48 & 12 & 39 & \star\\
12 & \star & 41 & 10 & \star & 43 & 4 & \star & 49 & \star & 13 & 40\\
39 & 13 & \star & 11 & \star & 44 & 5 & \star & 50 & 41 & \star & 14\\
\star & 43 & 17 & 36 & 12 & \star & 6 & \star & 51 & \star & 15 & 42\\
15 & \star & 44 & 37 & 13 & \star & 7 & \star & 52 & 43 & \star & 16\\
42 & 16 & \star & 38 & 14 & \star & 8 & \star & 53 & 17 & 44 & \star\\
\star & 46 & 20 & \star & 48 & 24 & 27 & 9 & \star & 18 & 45 & \star\\
18 & \star & 47 & \star & 49 & 25 & 28 & 10 & \star & \star & 19 & 46\\
45 & 19 & \star & \star & 50 & 26 & 29 & 11 & \star & 47 & \star & 20\\
\star & 49 & 23 & 18 & \star & 51 & 30 & 12 & \star & \star & 21 & 48\\
21 & \star & 50 & 19 & \star & 52 & 31 & 13 & \star & 49 & \star & 22\\
48 & 22 & \star & 20 & \star & 53 & 32 & 14 & \star & 23 & 50 & \star\\
\star & 52 & 26 & 45 & 21 & \star & 33 & 15 & \star & 51 & \star & 24 \\
24 & \star & 53 & 46 & 22 & \star & 34 & 16 & \star & 25 & 52 & \star \\
51 & 25 & \star & 47 & 23 & \star & 35 & 17 & \star & \star & 26 & 53
\end{bmatrix}\label{aq1}
\end{IEEEeqnarray}
\end{figure}
%

In the following we show the transitions of the $S$ matrix with $r$ on using our scheme (\textbf{Algorithm~\ref{algox2}} uses the circuits $\{1,2,3\}$ and $\{2,3,4\}$ for delivery).
\begin{IEEEeqnarray*}{l}
S = \begin{bmatrix}
1 & 1 & 1\\
2 & 2 & 2\\
2 & 2 & 2\\
1 & 1 & 1
\end{bmatrix} \xrightarrow{r = 18} 
\begin{bmatrix}
0 & 0 & 0\\
1 & 1 & 1\\
1 & 1 & 1\\
1 & 1 & 1
\end{bmatrix} \xrightarrow{r = 36}
\begin{bmatrix}
0 & 0 & 0\\
0 & 0 & 0\\
0 & 0 & 0\\
0 & 0 & 0
\end{bmatrix}.
\end{IEEEeqnarray*} 
It can be seen that our scheme achieves a rate $\frac{36}{9} = 4$  for a subpacketization level of $9$.
%

%

\section{Conclusion}\label{conclusion}
We constructed an SC-CC scheme that can accommodate the caches and the users within a subpacketization budget and showed that our scheme performs better than the existing schemes in some cases. We used a known method for constructing  designs from matrices to construct our scheme. Our original contribution in terms of methodology is a technique that uses the circuit structure of a matroid obtained from a matrix to construct the delivery phase. We also showed that our technique performs better than existing PDA based techniques in some cases. 


Our scheme assumes that the central server is oblivious to the user-to-cache association during placement. Over time, the central server may learn the expected number of users accessing each cache. Seemingly, such information may help reduce the average delay (or rate) further; the server may update the matrix (used to construct the  design) as per the learned distribution. In our future work, we plan to investigate how knowledge of some statistics of the user distribution may help further reduce delay in an SC-CC setup.


\appendices

\section{Proofs of Section~\ref{construct}}
\subsection{Proof of Lemma~\ref{lemmaRD}}\label{appendixRD}
\begin{IEEEproof}
%
Let the column vectors in $Q$ be denoted by $v_1,v_2,\ldots,v_{q^m}$. Since $Q$ contains every vector in $\mathbb{F}_q^m$ as a column, the set of vectors $v_1,v_2,\ldots,v_{q^m}$ forms a group under vector addition over $\mathbb{F}_q$. Let this group be denoted by $Q_G$. Let the row vectors of $G$ be denoted by $u_1,u_2,\ldots, u_k$. And let the dot product between a row vector $u$ and a column vector $v$ be denoted by $u.v$. 

For a row vector $u_i$ where $1\leq i\leq k$, we have $u_i.(v_{j_1} + v_{j_2}) = u_i.v_{j_1} + u_i.v_{j_2}$. So the map $f_i: \{v_1,v_2,\ldots,v_{q^m}\} \rightarrow \{0,1,\ldots,q-1\}$ given by $f(v_j) = u_i.v_j$ is a group homomorphism for $1\leq j\leq q^m$, $1\leq i\leq k$. Let the kernel of $f_i$ be denoted by $Ker(f_i)$ where $Ker(f_i) =\{ v_j \,|\, u_i.v_j = 0, 1\leq j\leq q^m \}$. We know that  $Ker(f_i)$ is a subgroup of $Q_G$ (see property 4, Theorem 10.1 of \cite{gallian} for proof). Homomorphism $f_i$ maps all elements of a coset of the kernel to the same element in $\{0,1,\ldots,q-1\}$, and members of distinct cosets are mapped to distinct elements in $\{0,1,\ldots,q-1\}$ (see property 5, Theorem 10.1 of \cite{gallian} for proof).  

Now, it can be seen that for any $e \in \{0,1,\ldots,q-1\}$, there is at least one column vector $v_j \in \{v_1,v_2,\ldots,v_{q^m}\}$, such that $f(v_j) = e$. For example, if $u_i[x] = g$ where $g \neq 0$, and if $v_j$ has $g^{-1}e$ in the $x^{\text{th}}$ row and $0$ in all other rows, then  $f(v_j) = e$.


Since there are $q$ elements in $\{0,1,\ldots,q-1\}$, there must be $q$ distinct left cosets of $Ker(f_i)$ (see property 5 and 6, Theorem 10.1 of \cite{gallian} for proof). Now, as each coset is of the same size (see page 145, property 7 of \cite{gallian}), each coset must have $q^{m-1}$ elements.  Then, from (\ref{def1}), it is immediate that all blocks contain $q^{m-1}$ elements (each block is nothing but the collection of all column indices of the column vectors contained in a coset). 


Since all blocks are equal-sized, $(X_G,\mathcal{A}_G)$ is a design. Let
\begin{IEEEeqnarray}{l}
\text{for } 1\leq i\leq k:\,\; \mathcal{P}_i = \{B(i,j) \,|\, 0\leq j\leq q-1\}.
\end{IEEEeqnarray}  
For a fixed $1\leq i\leq k$, since any element in $X_G$ is a column index, it is in $B(i,j)$ for some $0\leq j\leq q-1$, and so $\mathcal{P}_i$ partitions $X_G$ (sets $B(i,j)$ and $B(i,j^\prime)$ for $j\neq j^\prime$ are disjoint by definition). So $\mathcal{P}_i$ is a parallel class. 
\end{IEEEproof}

\subsection{Proof of Lemma~\ref{1lemma}}\label{appendix1lemma}
\begin{IEEEproof}
Let $u$ be a column vector of $Q$, and say $Gu = w$. For every vector $v \in N(G)$ (where $N(G)$ is the null space of $G$) we have $G(u + v) = w$. We know $N(G)$ has $q^{m -rank(G)}$ vectors. 

We now show $\{u + N(G)\}$ is the set of all vectors that multiply $G$ (on the right) to produce $w$. Say $y \notin \{u + N(G)\}$ and $Gy = w$. Then, $Gy = Gu$, or $G(y - u) = 0$, hence, $y \in \{N(G) + u\}$, which is a contradiction.

Since $D$ has $q^m$ columns, and any column vector of $D$ appears in $q^{m - rank(G)}$ distinct columns, exactly $q^{rank(G)}$ column vectors of $D$ must be distinct.
\end{IEEEproof}

\subsection{Proof of Corollary~\ref{1coro}}\label{appendix1coro}
\begin{IEEEproof}
From Lemma~\ref{1lemma} we know $D$ has $q^m$ distinct column vectors, each of which appears in only $1$ column. So every column vector in $\mathbb{F}_q^m$ appears in exactly one column of $D$. 

If $e_1,e_2 \in B(1,j_1) \cap B(2,j_2) \cap \cdots \cap B(m,j_m)$, then $D[i,e_1] = D[i,e_2] = j_i$ for $1\leq i\leq m$, and hence a column vector in which the $i^{\text{th}}$ row contains $j_i$ appears twice in $D$, which is a contradiction. So $B(1,j_1) \cap B(2,j_2) \cap \cdots \cap B(m,j_m)$ is a singleton set. This proves the first part of the Corollary.

Suppose $e = B(1,j_1) \cap B(2,j_2) \cap \cdots \cap B(m,j_m) = B(1,j_1^*) \cap B(2,j_2^*) \cap \cdots \cap B(m,j_m^*)$ where $\{j_1,j_2,\ldots,j_m\} \neq \{j_1^*,j_2^*,\ldots,j_m^*\}$. Then for some $1\leq x\leq m$ we must have $j_x \neq j_x^*$. However, since $e \in B(x,j_x)$ and $e \in B(x,j_x^*)$, we have $D[x,e] = j_x = j_x^*$, a contradiction. This proves the second part of the Corollary.

Alternatively, the second part of the corollary can also be proved using the pigeonhole principle. Since there are $q$ blocks in each parallel class, $m$ blocks from $m$ distinct parallel classes can be chosen in $q^m$ ways. So there exists $q^m$ sets $B(1,j_1) \cap B(2,j_2) \cap \cdots \cap B(m,j_m)$ for $0\leq j_1,j_2,\ldots,j_m\leq q-1$. 
Now, there exist $q^m$ elements in $\{1,2,\ldots,q^m\}$ and any element in it can be expressed as an intersection of $m$ blocks chosen from $m$ distinct parallel classes (each element in $\{1,2,\ldots,q^m\}$ must occur in exactly one block in each parallel class, as a parallel class partitions the set of points). Hence, if an element $e \in \{1,2,\ldots,q^m\}$ can be expressed as an intersection of two or more distinct set of $m$ blocks chosen from $m$ distinct parallel classes, then there must exist elements $e^\prime, e^* \in \{1,2\ldots,q^m\}$, $e^\prime \neq e^*$, such that  $e^\prime, e^* \in B(1,j_1) \cap B(2,j_2) \cap \cdots \cap B(m,j_m)$ for some $0\leq j_1,j_2,\ldots,j_m\leq q-1$, which is a contradiction to the first part of the lemma.
\end{IEEEproof}
\subsection{Proof of Corollary~\ref{2coro}}\label{appendix2coro}
\begin{IEEEproof}
As per Lemma~\ref{1lemma}, $D$ has $q^{m-1}$ distinct column vectors, each of which appears in $q$ distinct columns. Then a column vector that has $j_i$ in the $i^{\text{th}}$ row for $1\leq i\leq m$ appears in $q$ distinct columns. This proves the corollary.
\end{IEEEproof}
\subsection{Proof of Corollary~\ref{3coro}}\label{appendix3coro}
\begin{IEEEproof}
Since $\{1,2,\ldots,m+1\}$ is a circuit, we have $rank(G) = m$ and any $m$ rows are independent. As per Corollary~\ref{1coro}, $B(1,j_1) \cap B(2,j_2) \cap \cdots \cap B(m,j_m)$ is a singleton set. Say $e = B(1,j_1) \cap B(2,j_2) \cap \cdots \cap B(m,j_m)$. Let $l_{m+1} = D[m+1,e]$. Then $e \in B(m+1,l_{m+1})$. Since blocks of a parallel class are disjoint, $e \notin B(m+1,l)$ where $l \neq l_{m+1}$.
\end{IEEEproof}

\section{Proof of Theorem~\ref{thmx}}\label{proof:thmx}
In this section, we show the proof of Theorem~\ref{thmx}. In the proof, we will use several other auxiliary results, which we establish first in Sections~\ref{prop:C_y} and \ref{prop:J}.
\subsection{Properties of $C_y$ and $(X_G,\mathcal{A}_G)$}\label{prop:C_y}
As per Corollary~\ref{3coro}, the element $e_{\{(\alpha_1,l_1),(\alpha_2,l_2),\ldots,(\alpha_m,l_m)\}}$ belongs to only one block in the parallel class $\mathcal{P}_{\alpha_{m+1}}$. 
Due to Corollary~\ref{2coro}, $E_{\{(\alpha_1,l_1),(\alpha_2,l_2),\ldots,(\alpha_{i-1},l_{i-1}),(\alpha_{i+1},l_{i+1}),\ldots,(\alpha_m,l_m)\}}^{C_y}$ contains $q$ elements.
\begin{claim}\label{claim_apr16}
The $q$ elements contained in $E_{\{(\alpha_1,l_1),(\alpha_2,l_2),\ldots,(\alpha_{i-1},l_{i-1}),(\alpha_{i+1},l_{i+1}),\ldots,(\alpha_m,l_m)\}}^{C_y}$ belong to $q$ distinct blocks of the parallel class $\mathcal{P}_{\alpha_{m+1}}$, \textit{i.e.}, if $e_a,e_b \in E_{\{(\alpha_1,l_1),(\alpha_2,l_2),\ldots,(\alpha_{i-1},l_{i-1}),(\alpha_{i+1},l_{i+1}),\ldots,(\alpha_m,l_m)\}}^{C_y}$ and $e_a \in B(\alpha_{m+1},l_a)$ and $e_b \in B(\alpha_{m+1},l_b)$ then $l_a \neq l_b$.
\end{claim}
\begin{IEEEproof}
On the contrary, say there exists $e_a,e_b \in E_{\{(\alpha_1,l_1),(\alpha_2,l_2),\ldots,(\alpha_{i-1},l_{i-1}),(\alpha_{i+1},l_{i+1}),\ldots,(\alpha_m,l_m)\}}^{C_y}$ such that $e_a,e_b\in B(\alpha_{m+1},l)$ for some $0\leq l\leq q-1$. Then both $e_a$ and $e_b$ are contained in the set $E_{\{(\alpha_1,l_1),(\alpha_2,l_2),\ldots,(\alpha_{i-1},l_{i-1}),(\alpha_{i+1},l_{i+1}),\ldots,(\alpha_m,l_m),(\alpha_{m+1},l)\}}^{C_y}$, which must be a singleton set as per Corollary~\ref{1coro}. 
\end{IEEEproof}
The following claim can be proved proceeding similarly to Claim~\ref{claim_apr16}. 
\begin{claim}\label{claim_apr16b}
The $q$ elements contained in $E_{\{(\alpha_1,l_1),(\alpha_2,l_2),\ldots,(\alpha_{i-1},l_{i-1}),(\alpha_{i+1},l_{i+1}),\ldots,(\alpha_m,l_m)\}}^{C_y}$ belong to $q$ distinct blocks of the parallel class $\mathcal{P}_{\alpha_{i}}$ where $1\leq i\leq m$. 
\end{claim}

Since $Z_{(\alpha_i,l_i)}$ contains $t$ blocks, as per Claims~\ref{claim_apr16} and \ref{claim_apr16b}, $E_{\{(\alpha_1,l_1),(\alpha_2,l_2),\ldots,(\alpha_{i-1},l_{i-1}),(\alpha_{i+1},l_{i+1}),\ldots,(\alpha_m,l_m)\}}^{\{C_y, (\alpha_i,l_i)\}}$ contains $q-t$ elements, and since $E_{\{(\alpha_1,l_1),(\alpha_2,l_2),\ldots,(\alpha_{i-1},l_{i-1}),(\alpha_{i+1},l_{i+1}),\ldots,(\alpha_m,l_m)\}}^{\{C_y, (\alpha_i,l_i)\}}$ is a subset of $E_{\{(\alpha_1,l_1),(\alpha_2,l_2),\ldots,(\alpha_{i-1},l_{i-1}),(\alpha_{i+1},l_{i+1}),\ldots,(\alpha_m,l_m)\}}^{C_y}$, these $q-t$ elements must belong to $q-t$ distinct blocks in $\mathcal{P}_{\alpha_{m+1}}$ and $q-t$ distinct blocks in $\mathcal{P}_{\alpha_i}$.

\begin{claim}\label{claim2_apr16}
If $e \in E_{\{(\alpha_1,l_1),(\alpha_2,l_2),\ldots,(\alpha_{i-1},l_{i-1}),(\alpha_{i+1},l_{i+1}),\ldots,(\alpha_m,l_m)\}}^{\{C_y, (\alpha_i,l_i)\}}$ then $e \notin B(\alpha_{m+1},l_{m+1})$ where $e_{\{(\alpha_1,l_1),(\alpha_2,l_2),\ldots,(\alpha_m,l_m)\}} \in B(\alpha_{m+1},l_{m+1})$.
\end{claim}
\begin{IEEEproof}
Let $e \in E_{\{(\alpha_1,l_1),(\alpha_2,l_2),\ldots,(\alpha_{i-1},l_{i-1}),(\alpha_{i+1},l_{i+1}),\ldots,(\alpha_m,l_m)\}}^{\{C_y, (\alpha_i,l_i)\}}$ and $e \in B(\alpha_{m+1},l_{m+1})$. Then both $e$ and $e_{\{(\alpha_1,l_1),(\alpha_2,l_2),\ldots,(\alpha_m,l_m)\}}$ are contained in the set $B(\alpha_1,l_1) \cap B(\alpha_2,l_2)\cap \cdots \cap B(\alpha_{i-1},l_{i-1})\cap B(\alpha_{i+1},l_{i+1})\cap\cdots\cap B(\alpha_m,l_m)\cap B(\alpha_{m+1},l_{m+1})$, which contains only a single element as per Corollary~\ref{1coro}. Now, $e \neq e_{\{(\alpha_1,l_1),(\alpha_2,l_2),\ldots,(\alpha_m,l_m)\}}$ as $e$ does not belong to any block contained in $Z_{(\alpha_i,l_i)}$ and $e_{\{(\alpha_1,l_1),(\alpha_2,l_2),\ldots,(\alpha_m,l_m)\}} \in B(\alpha_i,l_i)$ where $B(\alpha_i,l_i) \in Z_{(\alpha_i,l_i)}$. So the claim is proved by contradiction.
\end{IEEEproof}

Hence, the $q-t$ element in $E_{\{(\alpha_1,l_1),(\alpha_2,l_2),\ldots,(\alpha_{i-1},l_{i-1}),(\alpha_{i+1},l_{i+1}),\ldots,(\alpha_m,l_m)\}}^{\{C_y, (\alpha_i,l_i)\}}$ must belong to $q-1$ distinct blocks in $\mathcal{P}_{\alpha_{m+1}} \setminus B(\alpha_{m+1},l_{m+1})$. 

\begin{claim}\label{claim3_apr16}
Let $P^*$ be a set of any $t$ blocks in $\mathcal{P}_{\alpha_{m+1}} \setminus B(\alpha_{m+1},l_{m+1})$ where $e_{\{(\alpha_1,l_1),(\alpha_2,l_2),\ldots,(\alpha_m,l_m)\}} \in B(\alpha_{m+1},l_{m+1})$. Then, there exists at least one $e^* \in E_{\{(\alpha_1,l_1),(\alpha_2,l_2),\ldots,(\alpha_{i-1},l_{i-1}),(\alpha_{i+1},l_{i+1}),\ldots,(\alpha_m,l_m)\}}^{\{C_y, (\alpha_i,l_i)\}}$ such that $e^*$ belongs to one of the blocks contained in $P^*$.
\end{claim}
\begin{IEEEproof}
Say none of the $q-t$ elements of $E_{\{(\alpha_1,l_1),(\alpha_2,l_2),\ldots,(\alpha_{i-1},l_{i-1}),(\alpha_{i+1},l_{i+1}),\ldots,(\alpha_m,l_m)\}}^{\{C_y, (\alpha_i,l_i)\}}$ belong to any of the blocks contained in $P^*$. Then the $q-t$ elements in \newline $E_{\{(\alpha_1,l_1),(\alpha_2,l_2),\ldots,(\alpha_{i-1},l_{i-1}),(\alpha_{i+1},l_{i+1}),\ldots,(\alpha_m,l_m)\}}^{\{C_y, (\alpha_i,l_i)\}}$ must belong to $q-t$ distinct blocks contained in $\mathcal{P}_{\alpha_{m+1}} \setminus \{B(\alpha_{m+1},l_{m+1})\cup P^*\}$. However $\mathcal{P}_{\alpha_{m+1}} \setminus \{B(\alpha_{m+1},l_{m+1})\cup P^*\}$ contains only $q-1-t$ blocks. Thereby, the claim is proved by showing a contradiction. 
\end{IEEEproof}
We have the following claim as a generalization of Claim~\ref{claim3_apr16}.
\begin{claim}\label{claim4_apr16}
Let $P^*$ be a set of any $t+k$ blocks in $\mathcal{P}_{\alpha_{m+1}} \setminus B(\alpha_{m+1},l_{m+1})$ where $k \leq q-t-1$,  $e_{\{(\alpha_1,l_1),(\alpha_2,l_2),\ldots,(\alpha_m,l_m)\}} \in B(\alpha_{m+1},l_{m+1})$. Then, there exists at least $k+1$ elements in  $E_{\{(\alpha_1,l_1),(\alpha_2,l_2),\ldots,(\alpha_{i-1},l_{i-1}),(\alpha_{i+1},l_{i+1}),\ldots,(\alpha_m,l_m)\}}^{\{C_y, (\alpha_i,l_i)\}}$ that belongs to $k+1$ distinct blocks contained in $P^*$.
\end{claim}
\begin{IEEEproof}
$\mathcal{P}_{\alpha_{m+1}} \setminus \{B(\alpha_{m+1},l_{m+1})\cup P^*\}$ has $q-1 -t-k$ blocks. So up to a maximum of  $q-1 -t-k$ elements of $E_{\{(\alpha_1,l_1),(\alpha_2,l_2),\ldots,(\alpha_{i-1},l_{i-1}),(\alpha_{i+1},l_{i+1}),\ldots,(\alpha_m,l_m)\}}^{\{C_y, (\alpha_i,l_i)\}}$ can belong to the $q-1 -t-k$ distinct blocks contained in $\mathcal{P}_{\alpha_{m+1}} \setminus \{B(\alpha_{m+1},l_{m+1})\cup P^*\}$ . So at least $(q-t) - (q-1-t-k) = k+1$ elements of $E_{\{(\alpha_1,l_1),(\alpha_2,l_2),\ldots,(\alpha_{i-1},l_{i-1}),(\alpha_{i+1},l_{i+1}),\ldots,(\alpha_m,l_m)\}}^{\{C_y, (\alpha_i,l_i)\}}$ must belong to the blocks contained in $P^*$.
\end{IEEEproof}

\subsection{Properties of $J_{\{(\alpha_1,l_1),(\alpha_2,l_2),\ldots,(\alpha_{i-1},l_{i-1}),(\alpha_{i+1},l_{i+1}),\ldots,(\alpha_m,l_m)\}}^{\{C_y,(\alpha_i,l_i)\}}$}\label{prop:J}
For this subsection we assume that $e_{\{(\alpha_1,l_1),(\alpha_2,l_2),\ldots,(\alpha_m,l_m)\}} \in B(\alpha_{m+1},l_{m+1})$.
\begin{claim}\label{may6a}
If $J_{\{(\alpha_1,l_1),(\alpha_2,l_2),\ldots,(\alpha_{i-1},l_{i-1}),(\alpha_{i+1},l_{i+1}),\ldots,(\alpha_m,l_m)\}}^{\{C_y,(\alpha_i,l_i)\}}[k] = l$ then for $1\leq w\leq k$ we have \newline $J_{\{(\alpha_1,l_1),(\alpha_2,l_2),\ldots,(\alpha_{i-1},l_{i-1}),(\alpha_{i+1},l_{i+1}),\ldots,(\alpha_m,l_m)\}}^{\{C_y,(\alpha_i,l_i)\}}[w] \in \{(l_{m+1}+1)_q, (l_{m+1}+2)_q, \ldots, (l_{m+1}+ (l - l_{m+1}))_q \}$. 
\end{claim}
\begin{IEEEproof}
The claim is immediate from \textbf{Algorithm~\ref{algox1}}. Since we have \newline $J_{\{(\alpha_1,l_1),(\alpha_2,l_2),\ldots,(\alpha_{i-1},l_{i-1}),(\alpha_{i+1},l_{i+1}),\ldots,(\alpha_m,l_m)\}}^{\{C_y,(\alpha_i,l_i)\}}[k] = l$, as per \textbf{Algorithm~\ref{algox1}} there exists $k$ elements in  $E_{\{(\alpha_1,l_1),(\alpha_2,l_2),\ldots,(\alpha_{i-1},l_{i-1}),(\alpha_{i+1},l_{i+1}),\ldots,(\alpha_m,l_m)\}}^{\{C_y, (\alpha_i,l_i)\}}$ that are contained in the blocks $B(\alpha_{m+1},y)$ where $y \in \{(l_{m+1}+1)_q, (l_{m+1}+2)_q, \ldots, (l_{m+1}+ (l - l_{m+1}))_q \}$. One may note, $(l_{m+1}+ (l - l_{m+1}))_q = l$ where $0\leq l\leq q-1$.
\end{IEEEproof}
The following claim is evident from \textbf{Algorithm~\ref{algox1}} and Claim~\ref{claim2_apr16}.
\begin{claim}\label{may6b}
If $J_{\{(\alpha_1,l_1),(\alpha_2,l_2),\ldots,(\alpha_{i-1},l_{i-1}),(\alpha_{i+1},l_{i+1}),\ldots,(\alpha_m,l_m)\}}^{\{C_y,(\alpha_i,l_i)\}}[k] = l$ then for $k\leq w\leq q-t$ we have $J_{\{(\alpha_1,l_1),(\alpha_2,l_2),\ldots,(\alpha_{i-1},l_{i-1}),(\alpha_{i+1},l_{i+1}),\ldots,(\alpha_m,l_m)\}}^{\{C_y,(\alpha_i,l_i)\}}[w] \in \mathcal{P}_{\alpha_{m+1}} \setminus \{\{ B(\alpha_{m+1},(l_{m+1}+1)_q), B(\alpha_{m+1},(l_{m+1}+2)_q), \ldots, B(\alpha_{m+1},(l_{m+1}+ (l -1 - l_{m+1}))_q)\} \cup B(\alpha_{m+1},l_{m+1}) \}$. 
\end{claim}

\begin{claim}\label{claim5_apr16}
$J_{\{(\alpha_1,l_1),(\alpha_2,l_2),\ldots,(\alpha_{i-1},l_{i-1}),(\alpha_{i+1},l_{i+1}),\ldots,(\alpha_m,l_m)\}}^{\{C_y,(\alpha_i,l_i)\}}[k] \in \{(l_{m+1} + k)_{q}, (l_{m+1} + k + 1)_{q}, \ldots, (l_{m+1} + k + t -1)_{q}\}$ for $1\leq k\leq q-t$.
\end{claim}
\begin{IEEEproof}
Let
\begin{IEEEeqnarray*}{l}
D_1 = \{(l_{m+1} + 1)_q, (l_{m+1} + 2)_q, \ldots, (l_{m+1} + k -1)_q\}\\
D_2 = \{(l_{m+1} + k)_{q}, (l_{m+1} + k + 1)_{q}, \ldots, (l_{m+1} + k + t -1)_{q}\}\\
D_3 = \mathcal{P}_{\alpha_{m+1}}\setminus\{D_1 \cup D_2 \cup B(\alpha_{m+1},l_{m+1})\}.
\end{IEEEeqnarray*}
It can be seen that $D_1$ contains $k-1$ elements, $D_2$ contains $t$ elements, and $D_3$ contains $q - (k+t)$ elements. We prove the claim by contradiction. 

Say $J_{\{(\alpha_1,l_1),(\alpha_2,l_2),\ldots,(\alpha_{i-1},l_{i-1}),(\alpha_{i+1},l_{i+1}),\ldots,(\alpha_m,l_m)\}}^{\{C_y,(\alpha_i,l_i)\}}[k]  \in D_1$. Then as per Claim~\ref{may6a} $k$ elements in $J_{\{(\alpha_1,l_1),(\alpha_2,l_2),\ldots,(\alpha_{i-1},l_{i-1}),(\alpha_{i+1},l_{i+1}),\ldots,(\alpha_m,l_m)\}}^{\{C_y,(\alpha_i,l_i)\}}[w] \,|\,1\leq w\leq k \}$ must be contained in $k-1$ elements in $D_1$, which is a contradiction.

Say $J_{\{(\alpha_1,l_1),(\alpha_2,l_2),\ldots,(\alpha_{i-1},l_{i-1}),(\alpha_{i+1},l_{i+1}),\ldots,(\alpha_m,l_m)\}}^{\{C_y,(\alpha_i,l_i)\}}[k] \in D_3$. Then as per Claim~\ref{may6b} $q-t-(k-1) = q-t-k+1$ elements in $J_{\{(\alpha_1,l_1),(\alpha_2,l_2),\ldots,(\alpha_{i-1},l_{i-1}),(\alpha_{i+1},l_{i+1}),\ldots,(\alpha_m,l_m)\}}^{\{C_y,(\alpha_i,l_i)\}}[w] \,|\,k\leq w\leq q-t \}$ must be contained in $q - t - k$ elements in $D_3$, which is a contradiction.

So it must be that $J_{\{(\alpha_1,l_1),(\alpha_2,l_2),\ldots,(\alpha_{i-1},l_{i-1}),(\alpha_{i+1},l_{i+1}),\ldots,(\alpha_m,l_m)\}}^{\{C_y,(\alpha_i,l_i)\}}[k] \in D_2$.
\end{IEEEproof}

\subsection{Proof of Theorem~\ref{thmx}}\label{prop:thm}
As in Section~\ref{examples}, in line~\algref{algox2}{line2} of \textbf{Algorithm~\ref{algox2}}, let $Y_{\{(\beta_1,l_1),(\beta_2,l_2),\ldots,(\beta_m,l_m)\}}^{\{j,r\}} = sum$ for the fixed values of $\beta_1,\beta_2,\ldots,\beta_m$, $l_1,l_2,\ldots,l_m$, $j$, and $r$ considered in the corresponding loop.
From line~\algref{algox2}{linex1} we know that
\begin{IEEEeqnarray*}{l}
a = B(\beta_1,l_1) \cap B(\beta_2,l_2) \cap \cdots \cap B(\beta_m,l_m).
\end{IEEEeqnarray*}
For $l^*$ defined in line~\algref{algox2}{line1} we have the following.
\begin{claim}
$e_{\{(\beta_1,l_1),(\beta_2,l_2),\ldots,(\beta_{i-1},l_{i-1}),(\beta_{i+1},l_{i+1}),\ldots,(\beta_m,l_m),(\beta_{m+1},l^*)\}}$ does not belong to any block contained in $Z_{(\beta_i,l_i)}$.
\end{claim}
\begin{IEEEproof}
By the definition of $J_{\{(\beta_1,l_1),(\beta_2,l_2),\ldots,(\beta_{i-1},l_{i-1}),(\beta_{i+1},l_{i+1}),\ldots,(\beta_m,l_m)\}}^{\{\bar{C},(\beta_i,l_i)\}}$ in \textbf{Algorithm~\ref{algox1}}, the block $B(\beta_{m+1},l^*)$ contains one element from $E_{\{(\beta_1,l_1),(\beta_2,l_2),\ldots,(\beta_{i-1},l_{i-1}),(\beta_{i+1},l_{i+1}),\ldots,(\beta_m,l_m)\}}^{\{\bar{C}, (\beta_i,l_i)\}}$. Using Corollary~\ref{1coro} it can be concluded that this element must be \newline $e_{\{(\beta_1,l_1),(\beta_2,l_2),\ldots,(\beta_{i-1},l_{i-1}),(\beta_{i+1},l_{i+1}),\ldots,(\beta_m,l_m),(\beta_{m+1},l^*)\}}$ and it is unique. And, by definition, no element in $E_{\{(\beta_1,l_1),(\beta_2,l_2),\ldots,(\beta_{i-1},l_{i-1}),(\beta_{i+1},l_{i+1}),\ldots,(\beta_m,l_m)\}}^{\{\bar{C}, (\beta_i,l_i)\}}$ is contained in $Z_{(\beta_i,l_i)}$.
%
\end{IEEEproof}
So any user accessing the cache $c_{(\beta_i,l_i)}$ cannot retrieve a subfile with index \newline $e_{\{(\beta_1,l_1),(\beta_2,l_2),\ldots,(\beta_{i-1},l_{i-1}),(\beta_{i+1},l_{i+1}),\ldots,(\beta_m,l_m),(\beta_{m+1},l^*)\}}$ from the cache.

\begin{claim}\label{apr20a}
For $1\leq i\leq m$, $W^{d(\beta_i,l_i,S[\beta_i,l_i+1])}_{e_{\{(\beta_1,l_1),(\beta_2,l_2),\ldots,(\beta_{i-1},l_{i-1}),(\beta_{i+1},l_{i+1}),\ldots,(\beta_m,l_m),(\beta_{m+1},l^*)\}}}$ is retrieved by user $u_{(\beta_i,l_i,S[\beta_i,l_i+1])}$ from $Y_{\{(\beta_1,l_1),(\beta_2,l_2),\ldots,(\beta_m,l_m)\}}^{\{j,r\}}$ in line~\algref{algox2}{line2} (line~\ref{line2} of \textbf{Algorithm~\ref{algox2}}).
\end{claim}
\begin{IEEEproof}
For $1\leq k\leq m$, $k \neq i$, cache $c_{(\beta_i,l_i)}$ stores the subfile (of all files) with index \newline $y_k = e_{\{(\beta_1,l_1),(\beta_2,l_2),\ldots,(\beta_{k-1},l_{k-1}),(\beta_{k+1},l_{k+1}),\ldots,(\beta_m,l_m),(\beta_{m+1},J_{\{(\beta_1,l_1),(\beta_2,l_2),\ldots,(\beta_{k-1},l_{k-1}),(\beta_{k+1},l_{k+1}),\ldots,(\beta_m,l_m)\}}^{\{\bar{C},(\beta_k,l_k)\}}[j])\}}$. So user $u_{(\beta_i,l_i,S[\beta_i,l_i+1])}$ knows  $W^{d(\beta_k,l_k,S[\beta_k,l_k+1])}_{y_k}$. Cache $c_{(\beta_i,l_i)}$ also stores the subfile (of all files) with index $e_{\{(\beta_1,l_1),(\beta_2,l_2),\ldots,(\beta_i,l_i),\ldots,(\beta_m,l_m)}$. So user $u_{(\beta_i,l_i,S[\beta_i,l_i+1])}$ knows $W^{d(\beta_{m+1},(l_{m+1}+j)_q,S[\beta_{m+1},(l_{m+1}+j)_q+1])}_{e_{\{(\beta_1,l_1),(\beta_2,l_2),\ldots,(\beta_m,l_m)\}}}$. 
%
\end{IEEEproof}

\begin{claim}
$e_{\{(\beta_1,l_1),(\beta_2,l_2),\ldots,(\beta_m,l_m)\}}$ does not belong to any block contained in $Z_{(\beta_{m+1},(l_{m+1} + j)_q)}$.
\end{claim}
\begin{IEEEproof}
We know $e_{\{(\beta_1,l_1),(\beta_2,l_2),\ldots,(\beta_m,l_m)\}} \in B(\beta_{m+1},l_{m+1})$ from line~\algref{algox2}{line3}. For $1\leq j\leq q-t$: $Z_{(\beta_{m+1}, (l_{m+1} + j)_q)} = \{B(\beta_{m+1},(l_{m+1} + j)_q), B(\beta_{m+1}, (l_{m+1} + j + 1)_q), \ldots, B(\beta_{m+1}, (l_{m+1} + j+t-1)_q)\}$.

If for some $0\leq w\leq t-1$ we have $B(\beta_{m+1},(l_{m+1} + j + w)_q) = B(\beta_{m+1},l_{m+1})$, then we have $(j + w)_q = 0$. Since $j \neq 0$, and as $j < q-t + 1$, or $j +t - 1 < q$, we know that $(j+w) \text{ mod } q \neq 0$. So $B(\beta_{m+1},l_{m+1}) \notin Z_{(\beta_{m+1}, (l_{m+1} + j)_q)}$. Then, from Corollary~\ref{3coro}, we can confirm the validity of the claim. 
\end{IEEEproof}
So any user accessing the cache $c_{(\beta_{m+1},(l_{m+1} + j)_q)}$ cannot retrieve a subfile with index \newline $e_{\{(\beta_1,l_1),(\beta_2,l_2),\ldots,(\beta_m,l_m)\}}$ from the cache it accesses.
\begin{claim}
$W^{d(\beta_{m+1},(l_{m+1} + j)_q),S[\beta_{m+1},(l_{m+1} + j)_q+1]}_{e_{\{(\beta_1,l_1),(\beta_2,l_2),\ldots,(\beta_m,l_m)\}}}$ is retrieved by user $u_{(\beta_{m+1},(l_{m+1} + j)_q,S[\beta_{m+1},(l_{m+1} + j)_q+1])}$ from $Y_{\{(\beta_1,l_1),(\beta_2,l_2),\ldots,(\beta_m,l_m)\}}^{\{j,r\}}$ in line~\algref{algox2}{line2}.
\end{claim}
\begin{IEEEproof}
We show that $e_{\{(\beta_1,l_1),(\beta_2,l_2),\ldots,(\beta_{i-1},l_{i-1}),(\beta_{i+1},l_{i+1}),\ldots,(\beta_m,l_m),(\beta_{m+1},l^*)\}}$ (where $l^*$ is defined in line~\algref{algox2}{line1}) belongs to one of the blocks is contained in $Z_{(\beta_{m+1},(l_{m+1} + j)_q)}$ for $1\leq i\leq m$. We achieve this by showing $B(\beta_{m+1},l^*) \in Z_{(\beta_{m+1},(l_{m+1} + j)_q)}$.

We know $Z_{(\beta_{m+1}, (l_{m+1} + j)_q)} = \{B(\beta_{m+1},(l_{m+1} + j)_q), B(\beta_{m+1}, (l_{m+1} + j + 1)_q), \ldots, B(\beta_{m+1}, (l_{m+1} + j+t-1)_q)\}$. From Claim \ref{claim5_apr16} we know $l^* \in  \{(l_{m+1} + j)_q, (l_{m+1} + j+1)_q, \ldots, (l_{m+1} + j+t-1)_q\}$ for $1\leq j\leq q-t$. Hence, $B(\beta_{m+1},l^*) \in Z_{(\beta_{m+1},(l_{m+1} + j)_q)}$.
\end{IEEEproof}

\begin{claim}\label{may7b}
After transmission of $Y_{\{(\beta_1,l_1),(\beta_2,l_2),\ldots,(\beta_m,l_m)\}}^{\{j,r\}}$ for $0\leq l_1,l_2,\ldots,l_{m} \leq q-1$, $0\leq j\leq q-t$, and all corresponding values of $r$, users $u_{(\beta_i,l_i,S[\beta_i,l_i+1])}$ for $1\leq i\leq m$ and \newline $u_{(\beta_{m+1},(l_{m+1} + j)_q), S[\beta_{m+1},(l_{m+1} + j)_q+1])}$ can retrieve their demanded files. Effectively,  user \newline $u_{(\beta_{m+1},l,S[\beta_{m+1},l+1])}$ for $0\leq l\leq q-1$ retrieves its demanded files.
\end{claim}
\begin{IEEEproof}
From Claim~\ref{claim_apr16} and \textbf{Algorithm~\ref{algox1}} it can be concluded that\newline $J_{\{(\beta_1,l_1),(\beta_2,l_2),\ldots,(\beta_{i-1},l_{i-1}),(\beta_{i+1},l_{i+1}),\ldots,(\beta_m,l_m)\}}^{\{\bar{C},(\beta_i,l_i)\}}[j] \neq J_{\{(\beta_1,l_1),(\beta_2,l_2),\ldots,(\beta_{i-1},l_{i-1}),(\beta_{i+1},l_{i+1}),\ldots,(\beta_m,l_m)\}}^{\{\bar{C},(\beta_i,l_i)\}}[j^*]$ for $1\leq j,j^* \leq q-t$, $j \neq j^*$ (as they belong to distinct blocks in $\mathcal{P}_{\beta_{m+1}}$). So, as per Claim~\ref{apr20a}, for $1\leq i\leq m$, $0\leq l_1,l_2,\ldots,l_{i-1},l_{i+1},\ldots,l_{m} \leq q-1$, $1\leq j\leq q-t$, user $u_{(\beta_i,l_i,S[\beta_i,l_i+1])}$ receives $(q-t)q^m$ distinct subfiles  $W^{d(\beta_i,l_i,S[\beta_i,l_i+1])}_{e_{\{(\beta_1,l_1),(\beta_2,l_2),\ldots,(\beta_{i-1},l_{i-1}),(\beta_{i+1},l_{i+1}),\ldots,(\beta_m,l_m),(\beta_{m+1},l^*)\}}}$. 

From $\{Y_{\{(\beta_1,l_1),(\beta_2,l_2),\ldots,(\beta_m,l_m)\}}^{\{j,r\}} \,|\, 1\leq j\leq q-t\}$, all users $u_{(\beta_{m+1},l,S[\beta_{m+1},l+1])}$ for all $0\leq l\leq q-1$ such that  $B(\beta_{m+1},l_{m+1}) \notin u_{(\beta_{m+1},l,S[\beta_{m+1},l+1])}$, receives the subfile with index \newline $e_{\{(\beta_1,l_1),(\beta_2,l_2),\ldots,(\beta_m,l_m)\}}$ of the file it demands. So after all transmissions (corresponding to the for loop in line~\algref{algox2}{linex2}) for $0 \leq l_1, l_2, \ldots, l_m \leq q-1$, all such users $u_{(\beta_{m+1},l,S[\beta_{m+1},l+1])}$ for $0\leq l\leq q-1$ would receive the file it demands.
%

To see the final part of the claim, first see all users in $U^{\bar{C}}$ where
\begin{IEEEeqnarray*}{l}
U^{\bar{C}} = \{u_{(\beta_{m+1},l,S[\beta_{m+1},l+1])}\,|\, 0\leq l\leq q-1\} \setminus \{u_{(\beta_{m+1},(l_{m+1} + j)_q), S[\beta_{m+1},(l_{m+1} + j)_q+1])}\,|\, 1\leq j\leq q-t \}
\end{IEEEeqnarray*} 
has the subfile indexed by $e_{\{(\beta_1,l_1),(\beta_2,l_2),\ldots,(\beta_m,l_m)\}}$ of all files. The reason is the following. See, $U^{\bar{C}}$ has $t$ users. A user $u_{(\beta_{m+1},l,S[\beta_{m+1},l+1])} \in U^{\bar{C}}$ for $l \in \{(l_{m+1} + q-t+1)_q, (l_{m+1} + q-t+2)_q,\ldots, l_{m+1}\}$ accesses the cache $Z_{(\beta_{m+1},l)} = \{B(\beta_{m+1},l), B(\beta_{m+1},(l+1)_q), \ldots, B(\beta_{m+1},(l+t-1)_q)$. Now, for $l = (l_{m+1} + q-t+1)_q + w$ where $0\leq w\leq t-1$, we have $B(\beta_{m+1},l_{m+1}) = B(\beta_{m+1},(l+t-1-w)_q)$; so $B(\beta_{m+1},l_{m+1}) \in Z_{(\beta_{m+1},l)}$. Now, we know $e_{\{(\beta_1,l_1),(\beta_2,l_2),\ldots,(\beta_m,l_m)\}} \in B(\beta_{m+1},l_{m+1})$ (line~\algref{algox2}{line3}).
\end{IEEEproof}
From Claim~\ref{may7b} we know the  users $u_{(\beta_i,l_i,S[\beta_i,l_i+1])}$ and $u_{(\beta_{m+1},l, S[\beta_{m+1},l+1])}$ for $1\leq i\leq m$, $0\leq l_1,l_2,\ldots,l_{m},l \leq q-1$ get their demanded files in the loops corresponding to the circuit $\bar{C}$. Before the next While loop, the $S$ matrix is updated to reflect the users that have been served.

In \textbf{Algorithm~\ref{algox2}}, $r$ keeps track of the total number of coded subfiles to be transmitted. The $flag$ variable ensures that $Y_{\{(\beta_1,l_1),(\beta_2,l_2),\ldots,(\beta_m,l_m)\}}^{\{j,r\}}$ gets transmitted only if at least one of the corresponding users gets benefited. 

Since any $1\leq i\leq n$ is a member of at least one $(m+1)$-length circuit, the demands of all users are delivered after the completion of \textbf{Algorithm~\ref{algox2}}. 

\end{document}